\documentclass[superscriptaddress,twocolumn,notitlepage,showkeys,showpacs,longbibliography]{revtex4-1}
\usepackage[utf8]{inputenc}
\usepackage{stmaryrd}
\usepackage{amsmath}
\usepackage{amssymb}
\usepackage{graphicx}
\usepackage{dcolumn}
\usepackage{bm}
\usepackage{multirow}
\usepackage[colorlinks,linkcolor=blue,citecolor=blue,urlcolor=blue]{hyperref}
\usepackage{color}
\usepackage{ulem}
\usepackage{comment}
\usepackage{booktabs}

\begin{document}
\title{
Three-dimensional higher-order saddle points induced flat bands in Co-based kagome metals
}
\author{Hengxin Tan}
\author{Yiyang Jiang}
\affiliation{Department of Condensed Matter Physics, Weizmann Institute of Science, Rehovot 7610001, Israel}

\author{Gregory T. McCandless}
\author{Julia Y. Chan}
\affiliation{Department of Chemistry and Biochemistry, Baylor University, Waco, TX 76798, USA}

\author{Binghai Yan}
\affiliation{Department of Condensed Matter Physics, Weizmann Institute of Science, Rehovot 7610001, Israel}

\begin{abstract}
The saddle point (van Hove singularity) exhibits a divergent density of states in 2D systems, leading to fascinating phenomena like strong correlations and unconventional superconductivity, yet it is seldom observed in 3D systems.
In this work, we have found two types of 3D higher-order saddle points (HOSPs) in emerging 3D kagome metals, YbCo$_6$Ge$_6$ and MgCo$_6$Ge$_6$. 
Both HOSPs exhibit a singularity in their density of states, which is significantly enhanced compared to the ordinary saddle point.
The HOSP near the Fermi energy generates a flat band extending a large area in the Brillouin zone, potentially amplifying the correlation effect and fostering electronic instabilities.
Two types of HOSPs exhibit distinct robustness upon element substitution and lattice distortions in these kagome compounds.
Our work paves the way for engineering exotic band structures, such as saddle points and flat bands, and exploring interesting phenomena in Co-based kagome materials.
\end{abstract}

\maketitle
\textit{Introduction}.
Kagome structures, noted for their two-dimensional (2D) corner-sharing triangular lattices, have captivated the condensed matter community \cite{Helton2007spin,liu2010spontaneous,xu2015intrinsic,nakatsuji2015large,nayak2016large,liu2018giant,ye2018massive,yin2018giant,wang2018large,liu2019magnetic,morali2019fermi,yin2019negative,kang2020dirac,yin2020quantum,liu2020orbital,Roychowdhury2022large,li2022manipulation,teng2022discovery,teng2023magnetism,tan2023prl,ma2021rare} due to their intricate crystal and electronic structures, which manifest in flat bands, Dirac points, and saddle points (SPs, also called van Hove singularity).
Amongst those characteristics, the SP sets the stage for various charge, spin, and pairing instabilities \cite{wen2010interaction,Kiesel2012sublattice,wang2013Competing,Kiesel2013unconventional,lin2021complex}.
The observation of charge density wave (CDW), pair density wave, and superconductivity in the quasi-2D kagome metals $A$V$_3$Sb$_5$ ($A=$ alkali metals) \cite{Ortiz2019new,yang2020giant,jiang2021unconventional,Ortiz2020Z2,yin2021superc,tan2021prl,chen2021roton,liang2021three,zhao2021cascade,Li2021observation,Chen2021double,fu2021quantum,feng2021chiral,wang2021electronic,wu2021nature,denner2021analysis} in close relation to the SP marks a pivotal advancement in this domain.
The SP has also been closely connected to the unconventional superconductivity and correlated insulator in twisted bilayer graphene \cite{choi2019electronic,kerelsky2019maximized,liu2019magnetism}.

The divergent density of states (DOS) of the SP in the 2D kagome lattice is at the heart of the rich physics induced by the SP. However, in three-dimensional (3D) momentum space, the SP does not lead to a DOS divergence, decreasing the expectation of electronic instabilities and correlation phenomena from the SP. 
One realistic and promising strategy to elevate the DOS of 3D SP could be to flatten its energy dispersion by the higher-order saddle point (HOSP) \cite{yuan2019magic,Isobe2019supermetal,Classen2020competing}, which is characterized by a higher-order momentum-dependent energy dispersion.
Notably, the SP's energy dispersion transforms into a flat band along the flattening direction at the flattening limit.
For 2D cases, the HOSP leads to a power-law divergent DOS \cite{yuan2019magic,Isobe2019supermetal}, compared to the logarithmically divergent DOS of the ordinary SP with a quadratic momentum-dependent dispersion, which results in complex phases, such as the supermetal, density waves, superconductivity, Pomeranchuk instability, and Lifshitz transition \cite{Classen2020competing,Isobe2019supermetal,guerci2022higher,han2023enhanced,Castro2023emergence,zervou2023fate}.
In 3D systems, the enhanced DOS of a HOSP indicates a stronger tendency toward these intriguing properties that are less expected from the ordinary 3D SP.

In addition, in most realistic kagome materials, flat bands and Dirac points are away from the Fermi energy; SPs are close to the Fermi energy \cite{ye2018massive,tan2021prl,ma2021rare,hu2022tunable,tan2023prl,teng2023magnetism,peng2021realizing,kang2020dirac}. The energy separation between these key band features minimizes their interaction, diluting the potential for emergent quantum phenomena from their intertwining.
SPs are generally connected to Dirac points in kagome materials. Flattening SPs' dispersion by HOSP naturally brings Dirac points energetically closer to SPs, which enables the study of SP, flat band, and Dirac physics at the same energy level \cite{Abhay2023visualizing}.
For example, a recent semiclassical theory \cite{peshcherenko2024sublinear} suggests the vicinity of Dirac points to SPs leads to the sublinear temperature-dependent transport properties observed in kagome materials \cite{mozaffari2023universal,ye2024hopping}.
From these perspectives, HOSPs in 3D materials are intriguing and promising. However, a lack of material realization for 3D HOSPs limits relevant discussions.

In this study, we discuss the distinctive DOS behavior of 3D HOSPs and propose YbCo$_6$Ge$_6$ and MgCo$_6$Ge$_6$ as promising candidates to realize 3D HOSPs.
These kagome metals exhibit two types of HOSPs ($\zeta_1$ and $\zeta_2$) with dominant quartic momentum-dependent energy dispersions at the Brillouin zone boundary.
The mixed-type HOSP $\zeta_1$ (at $-$0.017 eV) engenders a flat band along the $L-H$ line due to the small quartic term coefficient in its dispersion, drawing the connected Dirac point close to the Fermi energy.
The pure-type HOSP $\zeta_2$ (at $-$0.15 eV) generates a shorter flat band due to its larger quartic term coefficient.
$\zeta_1$ is sensitive to Yb-site element substitution but robust against the prevalent lattice distortion$-$Co triangle twisting$-$in these Co-based kagome materials.
In contrast, the $\zeta_2$ exhibits sensitivity to lattice distortion but is resistant to element substitution.
These results suggest the robustness of $\zeta_1$-related properties (e.g., Pomeranchuk instability and electronic nematicity) against temperature, and element substitution can be a useful knob for their manipulation.

\begin{figure}[tbp]
\includegraphics[width=0.99\linewidth]{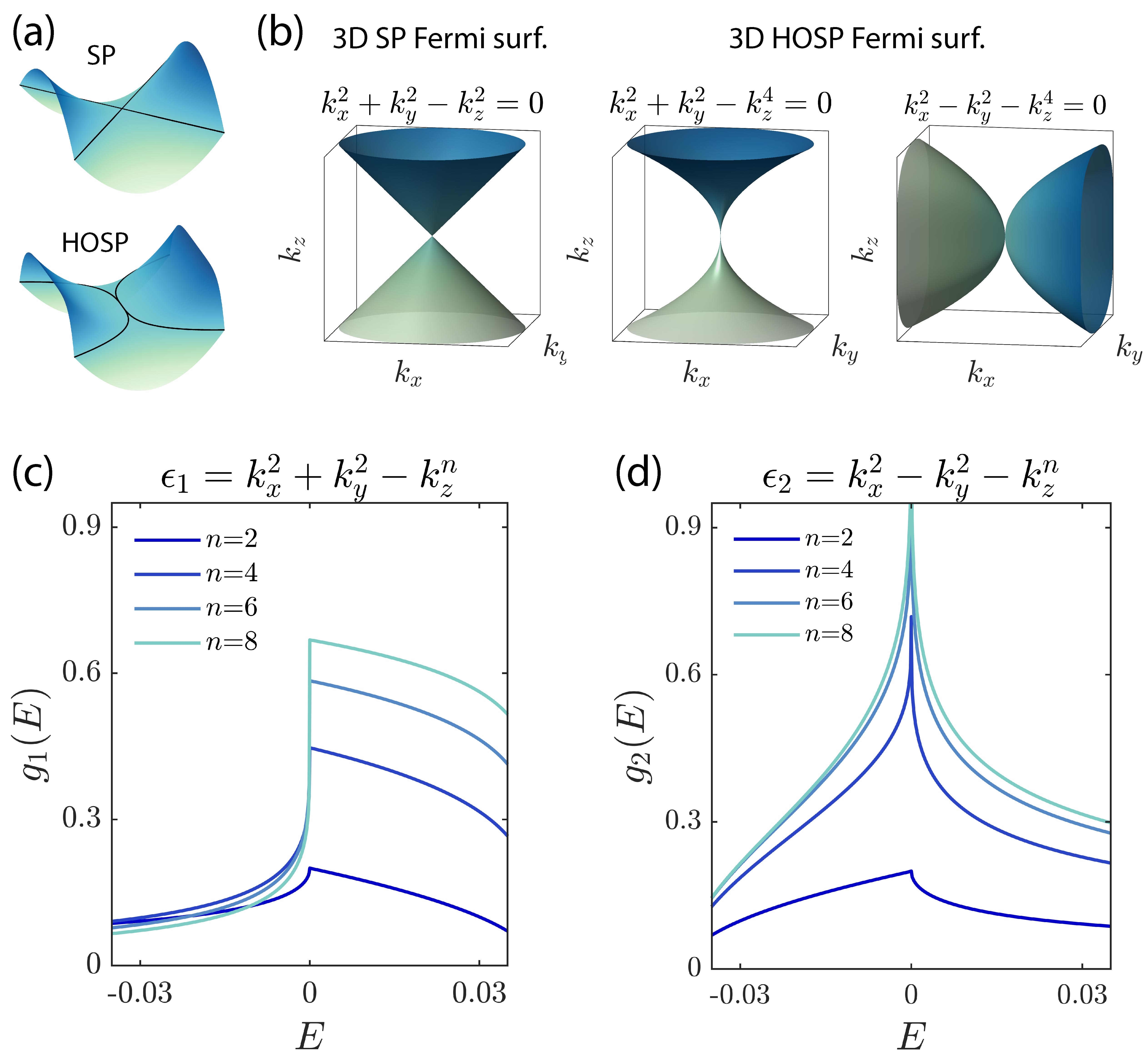}
\caption{\label{Fig1} (a) Comparison between an ordinary 2D SP and HOSP, with black lines and curves illustrating the energy contour at the SP energy (i.e., Fermi surface).
(b) Fermi surfaces of the ordinary 3D SP compared to 3D HOSPs.
(c) and (d), DOS [$g(E)$] of the two different 3D HOSPs in Eq. \eqref{eq1}, (c) for $\epsilon_1$ and (d) for $\epsilon_2$, plotted against $n$. A truncation radius of 0.2 for momenta $k$ is employed in the integral (Supplementary section II \cite{SM}). The DOS is scaled by $(4\pi^2)^{-1}$. All coefficients $c_{i,n}$ in (b)-(d) have been set to one.}
\end{figure}

\textit{Analysis of 3D HOSP.}
The energy dispersion of a 3D SP, noted as $\epsilon$ (to distinguish from its material realization $\zeta$), is given by $\epsilon(\textbf{k}) = \sum_{i,n} c_{i,n} k_i^{n}$, where $i$ represents the spatial directions \{$x, y, z$\}; $n$ are integers larger than one. The SP's energy is zero energy; momentum \textbf{k} is measured relative to the SP's momentum. Coefficients $c_{i,n}$ along three directions vary in sign to form an SP. $n$ greater than two in any direction categorizes the SP as a HOSP. Considering the mirror symmetries of the SP in our material below, we constrain $n$ to be even in all our discussions. Without loss of generality, the simplest HOSP might take the following two forms,
\begin{equation}
\begin{cases}
\epsilon_1(\textbf{k}) = c_x k_x^2 + c_y k_y^2 - c_{z} k_z^{n},\\
\epsilon_2(\textbf{k}) = c_x k_x^2 - c_y k_y^2 - c_{z} k_z^{n},\\
\end{cases}
\label{eq1}
\end{equation}
where we have taken all coefficients positive and the higher-order on $k_z$.
Fermi surfaces of two HOSPs show a tangential touching at the SP's momentum, compared to the Dirac cone-like (linear) touching of the ordinary SP ($n=2$) in Fig. \ref{Fig1}(a)\&(b) (see also Supplementary Fig. S2 \cite{SM}).

The DOS, $g(E)$, in the vicinity of two HOSPs, can be derived (Supplementary section II \cite{SM}) and depicted in Fig. \ref{Fig1}(c)\&(d). It shows that under a finite $c_z$, $g(E)$ at zero energy increases rapidly with an increasing $n$ because a higher $n$ flattens the energy dispersion around the SP.
While $\epsilon_1$ and $\epsilon_2$ have the same DOS with $n=2$ (bear a sign reverse in energy), their DOS behaves differently when $n>2$.
For $\epsilon_1$, the DOS is finite, and its first derivative is power-law divergent with $|E|^{\frac{1}{n} - 1}$ when $E \rightarrow 0^-$. On the other hand, the DOS and its first derivative are both finite when $E \rightarrow 0^+$, as shown in Fig. \ref{Fig1}(c).
For $\epsilon_2$ in Fig. \ref{Fig1}(d), however, the DOS's first derivative becomes power-law divergent with $|E|^{\frac{1}{n} - 1}$ when $E$ approaches zero from both sides (see also Ref. \cite{yuan2020class}), resulting in a pronounced but finite DOS peak.
We mention that the coefficient $c_z$ scales the DOS with a factor $(\sqrt[n]{c_z})^{-1}$ \cite{SM}. A $c_z$ value of zero implies a divergent DOS due to the perfect flat band along $k_z$. 
These observations underscore that a larger $n$ and smaller $c_z$ promote a flatter dispersion and higher DOS of the 3D SP, potentially amplifying electronic instabilities and correlation effects of a 3D SP \cite{Classen2020competing,Isobe2019supermetal,guerci2022higher,han2023enhanced,Castro2023emergence,zervou2023fate}.

\begin{figure*}[tbp]
\includegraphics[width=0.8\linewidth]{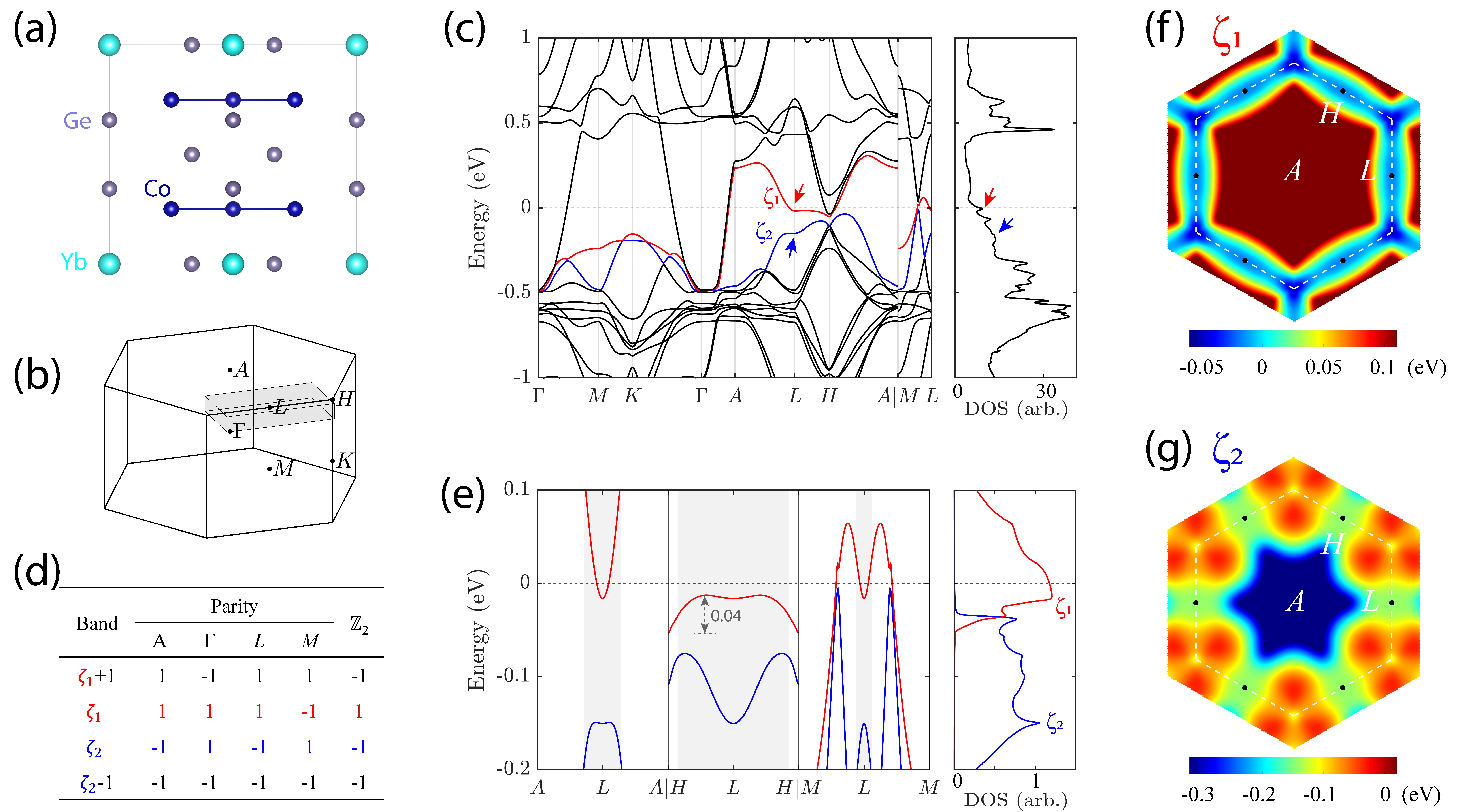}
\caption{\label{Fig2} (a) Crystal structure of YbCo$_6$Ge$_6$.
(b) High-symmetry points in the Brillouin zone.
(c) Band structure and DOS of YbCo$_6$Ge$_6$ incorporating spin-orbit coupling from DFT. Notice that all bands are doubly degenerate. HOSPs $\zeta_1$ and $\zeta_2$ are labeled.
(d) Parity and $\mathbb{Z}_2$ index of $\zeta_1$, $\zeta_2$ bands and one band below $\zeta_2$ (i.e., $\zeta_2-1$), one above $\zeta_1$ ($\zeta_1+1$).
(e) A zoom-in of the band dispersion of two HOSPs along three perpendicular directions ($k$-path length has been renormalized). The right panel shows the DOS contributed separately by two HOSPs [the integral volume for DOS is shown by the grey area in (e) and (b)].
(f) Momentum distribution of the $\zeta_1$ band on the $A-L-H$ plane [the red band in (c)]. The white hexagons in (f) and (g) denote the Brillouin zone boundary.
(g) Analogous to (f) but showcasing the $\zeta_2$ band [the blue band in (c)].
}
\end{figure*}

\textit{Material realization of 3D HOSPs.}
Here, we present the material realization of HOSPs near the Fermi level in non-magnetic kagome metals YbCo$_6$Ge$_6$ \cite{dzyanyi1995crystal,Fedyna1999magnetic} and MgCo$_6$Ge$_6$ \cite{Gieck2006synthesis}. 
YbCo$_6$Ge$_6$ exhibits disorder in experiment \cite{weiland2020refine} (named Yb$_{0.5}$Co$_3$Ge$_3$ therein). However, the HfFe$_6$Ge$_6$-type ordered lattice in Fig. \ref{Fig2}(a) represents a reasonable refinement for the disordered structure.
Notably, MgCo$_6$Ge$_6$ shows a well-ordered HfFe$_6$Ge$_6$-type structure \cite{Gieck2006synthesis}.
Given the property parallels between YbCo$_6$Ge$_6$ and MgCo$_6$Ge$_6$ (space group $P6/mmm$), our analysis primarily centers on the former, noting any pertinent distinctions.

The density functional theory (DFT)-derived electronic structure of YbCo$_6$Ge$_6$ is displayed in Fig. \ref{Fig2}(c), which showcases a $\mathbb{Z}_2$ topological metal, as unveiled by parity analysis at time-reversal invariant momenta \cite{Fu2007topological} in Fig. \ref{Fig2}(d). The non-trivial topology promises topological surface states, which might be buried in bulk bands due to their presence at the Fermi energy (Supplementary Fig. S6 \cite{SM}). Except for bundles of flat bands at around $\pm$0.5 eV that are primarily associated with Co $d$ orbitals, the low-energy electronic structure is featured with two pronounced HOSPs at $L$, $\zeta_1$ at $-$0.017 eV and $\zeta_2$ at $-$0.150 eV, alongside adjacent Dirac points at the $H$ point (gapped out by spin-orbit coupling). By polynomial fitting the DFT band dispersion along three orthogonal directions ($L-A$ as $x$, $L-H$ as $y$, and $L-M$ as $z$), energy expressions near two HOSPs are,
\begin{equation}
\begin{cases}
    \zeta_1(\textbf{k}) = -0.017 + 3.7 k_x^2 + 23.5 k_z^2 - 1.3 k_y^4\\
    \zeta_2(\textbf{k}) = -0.150 + 2.2 k_y^2 - 34.7 k_z^2 - 30.6 k_x^4\\
    \end{cases}
    \label{eq2}
\end{equation}
where \textbf{k}=($k_x$,$k_y$,$k_z$) is measured from the $L$ point, with energy in eV. Mirror symmetries of $L$ point guarantee the absence of any odd-order \textbf{k} dependence in these expressions. While $\zeta_1$ inherently exhibits a quadratic dependence on $k_y$ and $\zeta_2$ on $k_x$, their coefficients are around 0.1 and thus considered negligible in Eq. \ref{eq2} (a justification is presented in Supplementary Fig. S3 \cite{SM}). Notice that $\zeta_1$ and $\zeta_2$ have similar format as $\epsilon_1$ and $\epsilon_2$ in Eq. \eqref{eq1} with $n=4$.

A closer look at the dispersion of two HOSPs along three perpendicular directions is shown in Fig. \ref{Fig2}(e).
Due to the small coefficient ($-$1.3 eV$\cdot$\AA$^4$) of the quartic term, $\zeta_1$ induces a nearly flat band along the $L-H$ line (the eigenvalue difference between $L$ and $H$ is about 0.04 eV), which covers the hexagonal Brillouin zone boundary (on $A-L-H$ plane) as depicted by its momentum distribution in Fig. \ref{Fig2}(f). The DOS from this HOSP $\zeta_1$ [right panel of Fig. \ref{Fig2}(e)] is similar to that of $\epsilon_1$ in Fig. \ref{Fig1}(c).
Because this $\zeta_1$-associated band constitutes even flatter dispersions at the Fermi energy away from the $L-H$ line (Supplementary Fig. S5 \cite{SM}), it constitutes a total DOS peak closer to the Fermi energy in Fig. \ref{Fig2}(c).
$\zeta_1$ represents a mixed-type SP \cite{Kiesel2012sublattice,wu2021nature}, primarily contributed by all three Co atoms ($d_{z^2}$ and $d_{xz}$/$d_{yz}$ orbitals) in each kagome layer (Supplementary Fig. S4 \cite{SM}). This band contributes one branch to the gapped Dirac point at $H$ at around $-$0.06 eV, highlighting the potential interaction between HOSP-induced properties and Dirac band topology.

The HOSP $\zeta_2$ also generates a short flat dispersion along the $A-L$ line. But due to a much larger coefficient of the quartic term ($-$30.6 eV$\cdot$\AA$^4$), it disperses prominently away from $L$. 
The DOS behavior of this HOSP in Fig. \ref{Fig2}(e) is similar to that of $\epsilon_2$ in Fig. \ref{Fig1}(d), in line with their similar expressions.
The $\zeta_2$-associated band contributes a convex in the total DOS [Fig. \ref{Fig2}(c)], where the sharp peak from the HOSP $\zeta_2$ is hidden.
The momentum distribution of the $\zeta_2$ band is presented in Fig. \ref{Fig2}(g). The proximity of $\zeta_2$ to the Dirac point at $H$ in momentum and energy domains also suggests potential mutual influences.
Notably, $\zeta_2$ is a pure-type SP, primarily contributed by $d_{xy}$/$d_{x^2-y^2}$ orbitals of a single Co atom in each kagome sublattice (Supplementary Fig. S4 \cite{SM}).

\begin{figure}[tbp]
\includegraphics[width=0.95\linewidth]{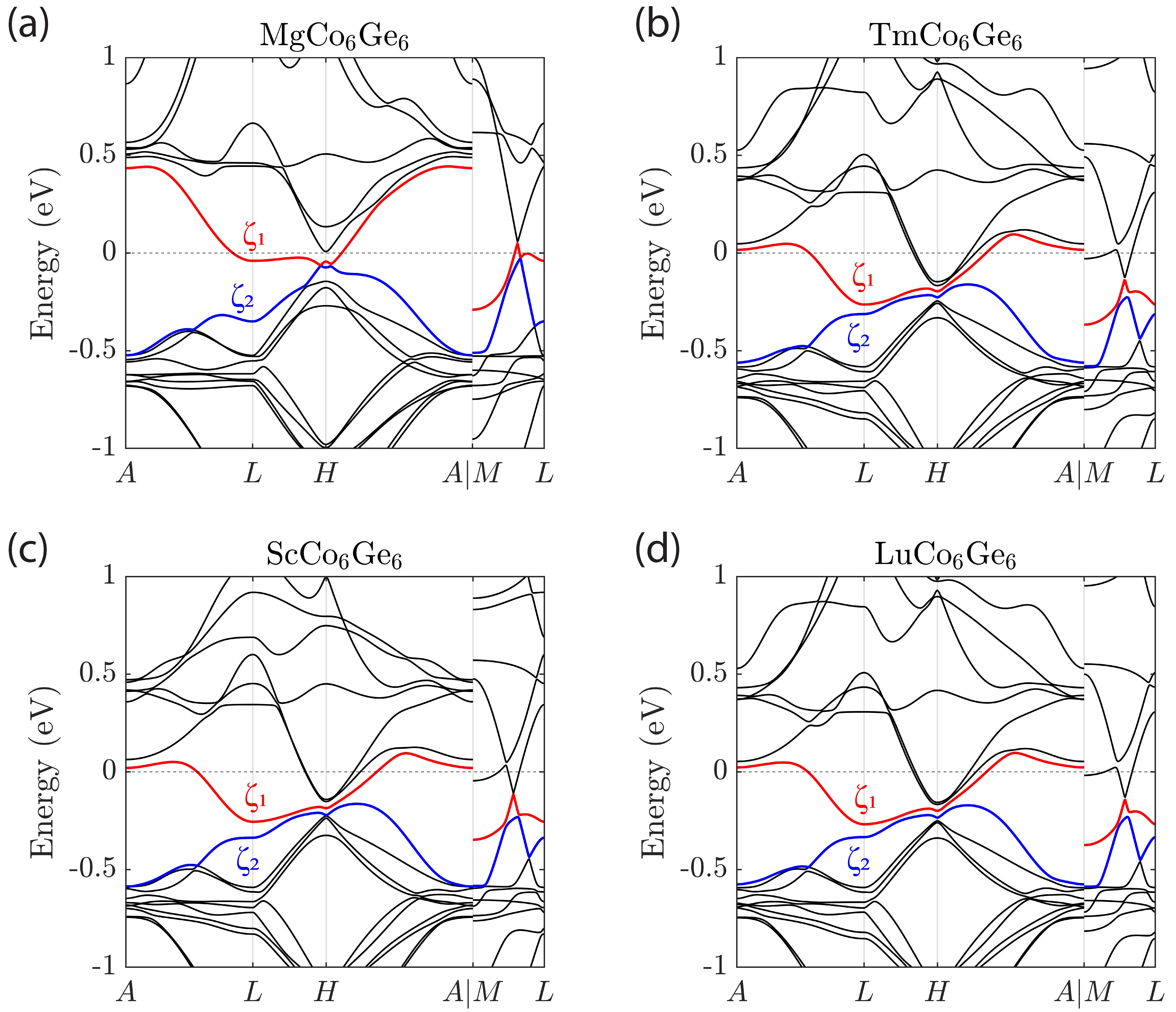}
\caption{\label{Fig3} Band structure of (a) MgCo$_6$Ge$_6$, (b) TmCo$_6$Ge$_6$, (c) ScCo$_6$Ge$_6$, and (d) LuCo$_6$Ge$_6$ incorporating spin-orbit coupling. For the iso-valence substitution (MgCo$_6$Ge$_6$), $\zeta_1$ remains a HOSP while $\zeta_2$ becomes an ordinary SP. For other compounds, $\zeta_1$ becomes a 3D parabola while $\zeta_2$ remains a HOSP.
}
\end{figure}

\textit{Element substitution effect.}
In the structurally analogous MgCo$_6$Ge$_6$, where Mg has the same valence electrons as Yb, the $\zeta_1$ and its consequent flat band are also present near the Fermi energy [Fig. \ref{Fig3}(a)]. The $\zeta_2$ manifests as an ordinary SP without a conspicuous flat band. In carrier-doped variants where Lu, Tm, or Sc substitute Yb [Fig. \ref{Fig3}(b-d)], $\zeta_2$ consistently appears as a HOSP; the $\zeta_1$ band along $L-H$ becomes an upward parabola, transforming $\zeta_1$ into a 3D parabola that opens upward in all directions. Additionally, $\zeta_1$ and $\zeta_2$ shift deeper in energy. This transition, marked by the loss of the HOSP $\zeta_1$ and its associated flat band, may significantly diminish the electron instability and correlation effects potentially observed in YbCo$_6$Ge$_6$ and MgCo$_6$Ge$_6$. Thus, the latter compounds hold more promise for exploring HOSP-driven phenomena. These materials, including their element-substituted variants, offer a rich platform to study the impact of carrier doping on HOSP attributes.

\begin{figure}[tbp]
\includegraphics[width=0.9\linewidth]{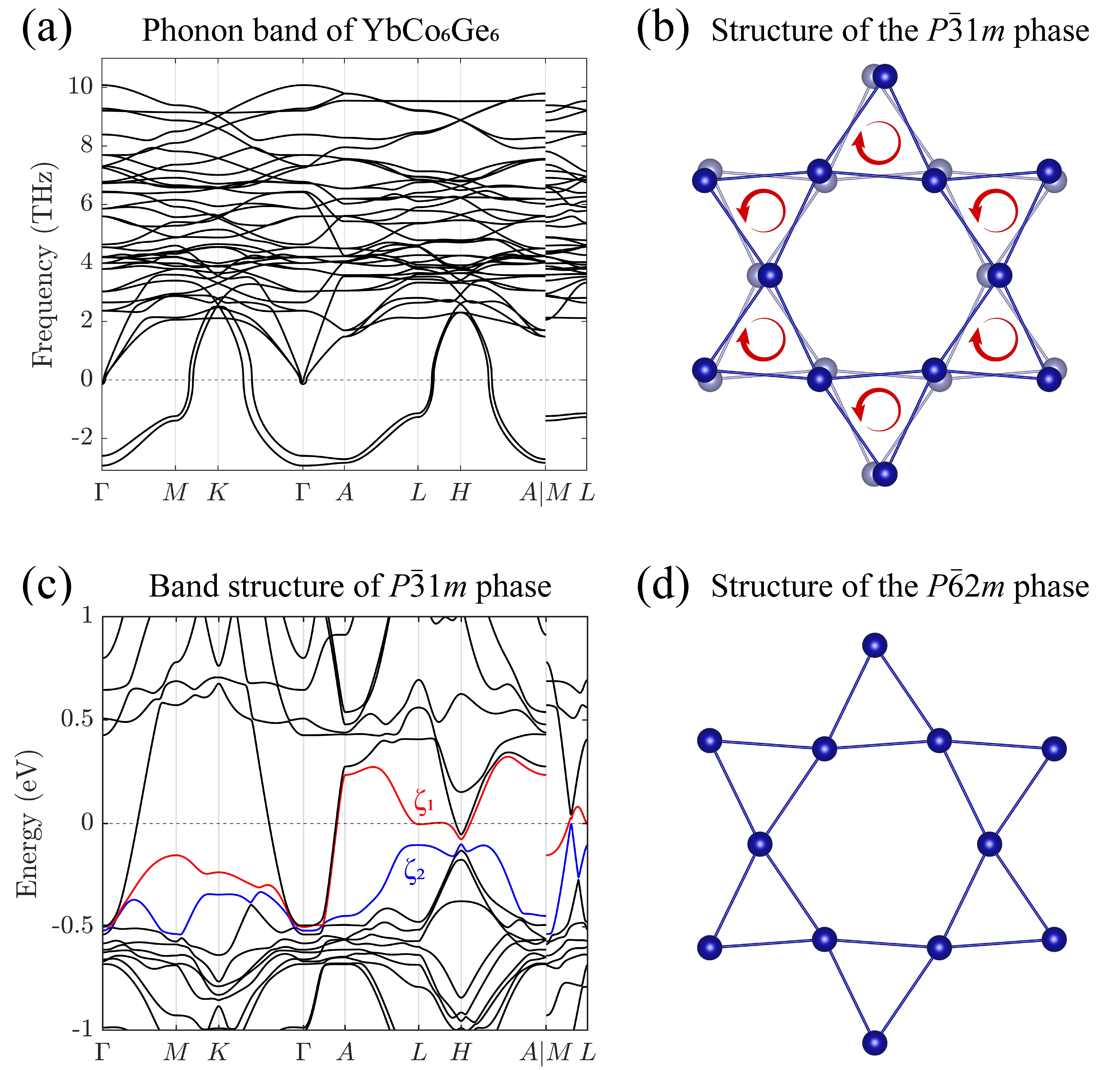}
\caption{\label{Fig4} (a) Phonon band structure of YbCo$_6$Ge$_6$ ($P6/mmm$ phase). Imaginary frequency is depicted as negative values.
(b) The $P\bar{3}1m$ structure displays only Co kagome sublattices for clarity (blue for the upper layer and grey for the lower layer). The red circular arrows show the relative local rotation of the upper to the lower Co-triangles.
(c) Band structure of the $P\bar{3}1m$ phase with spin-orbit coupling. The HOSP $\zeta_1$ and the transformed $\zeta_2$ (now a downward-opening 3D parabola) are highlighted. (d) The top view of the $P\bar{6}2m$ structure where two kagome layers are overlaid and totally eclipse each other.
}
\end{figure}

\textit{Effect of lattice distortion.}
Lattice dynamics analysis of the compounds reveals two prominent imaginary phonon modes at the $\Gamma$ point, exemplified by the phonon dispersion in YbCo$_6$Ge$_6$ [Fig. \ref{Fig4}(a)]. These modes correspond to in-phase and out-of-phase rotations of Co-triangles within the dual kagome layers of each unit cell. While these rotations maintain the translational symmetry, hence not prompting a CDW, they transition the lattice's six-fold rotational symmetry to three-fold and lead to two dynamically stable and energetically favorable structures$-$one ($P\bar{3}1m$, No. 162) with total energy $-$40 meV/unit-cell relative to the pristine structure, and the other ($P\bar{6}2m$, No. 189) $-$27 meV. The $P\bar{3}1m$ structure features out-of-phase Co-triangle rotation between adjacent kagome layers [Fig. \ref{Fig4}(b)], which aligns well with experimental findings for YbCo$_6$Ge$_6$ \cite{wang2022electronic}.
This distortion might offer potential avenues for local chiral transport studies despite the absence of global layer twisting.
Conversely, synchronous Co-triangle rotation of both kagome layers results in the noncentrosymmetric $P\bar{6}2m$ structure [Fig. \ref{Fig4}(d)]. It's noteworthy that inversion symmetry breaking is atypical in known CDW kagome materials \cite{tan2021prl,Arachchige2022charge,teng2022discovery,tan2023prl,tan2023emergent,kang2023charge}. The advantages of breaking inversion symmetry are significant, particularly evident in phenomena like Rashba-type spin-orbit coupling and non-linear transport.
Figure \ref{Fig4}(a) also features two imaginary phonon modes at the $A$ point, leading to a 1$\times$1$\times$2 CDW (space group $P6_3/mcm$, No. 193), which involves alternative stacking of $P\bar{3}1m$ and $P\bar{6}2m$ phases. Notably, while the $P\bar{3}1m$ has the lowest total energy in YbCo$_6$Ge$_6$, the CDW phase ($P6_3/mcm$) is the ground state in MgCo$_6$Ge$_6$ (Supplementary Fig. S8 \cite{SM}), which is confirmed by a recent experiment \cite{Sinha2021twisting}.

The band structure of the $P\bar{3}1m$ phase is depicted in Fig. \ref{Fig4}(c), exhibiting minimal band structure alterations near the Fermi energy, with the notable exception of $\zeta_2$, which transitions from a HOSP to a downward-opening 3D parabola. This modification primarily stems from its in-plane $d$ orbital ($d_{xy}$/$d_{x^2-y^2}$) features, which are the most influenced by the in-plane Co-triangle rotation. $\zeta_1$, predominantly derived from the out-of-plane orbitals ($d_{z^2}$, $d_{xz}$/$d_{yz}$), retains its HOSP characteristics, preserving the associated flat band along the $L-H$ line. A similar impact of structural distortion on $\zeta_1$ and $\zeta_2$ is observed in the $P\bar{6}2m$ phase and the CDW phase $P6_3/mcm$ (Supplementary Fig. S7 and Fig. S8 \cite{SM}).

\textit{Discussions.} Given the clarity of the band structure of these Co-based materials near the Fermi energy, it's plausible that the two HOSPs and associated Dirac points predominantly influence the electronic properties near the Fermi energy.
The enhanced DOS of these HOSPs is promising to facilitate the orbital-correlation-driven electronic nematicity observed in kagome superconductor CsTi$_3$Bi$_5$ \cite{li2023electronic}.
The HOSP has recently been observed to induce the Pomeranchuk instability in kagome magnet Co$_3$Sn$_2$S$_2$ \cite{Pranab}. 
Thus, we speculate such effects might also be present in these Co-based kagome metals, which call for experimental verification.
The robustness of the HOSP $\zeta_1$ against structural distortions in YbCo$_6$Ge$_6$ and MgCo$_6$Ge$_6$, in contrast to its sensitivity to chemical doping, may indicate the resistance of such HOSP-induced phenomena to temperature variations.

A semiclassical theory \cite{peshcherenko2024sublinear} suggests the presence of both Dirac point and SP near the Fermi energy, which has been the case in Co-based kagome metals, would result in the non-Fermi-liquid-like transport behavior.
Indeed, recent experiments revealed a sublinear temperature-dependent resistivity in YbCo$_6$Ge$_6$ in the temperature range of (75, 200) K \cite{weiland2020refine,wang2022electronic} (similar behavior of MgCo$_6$Ge$_6$ at (75, 250) K in Ref. \cite{Sinha2021twisting}, see Fig. S10 \cite{SM} for details), which may signify the existence of HOSPs.
Most importantly, Ref. \cite{Sinha2021twisting} reveals the negligible response of resistivity and specific heat to the CDW transition in MgCo$_6$Ge$_6$, potentially demonstrating the predicted robustness of $\zeta_1$-dominated properties against temperature.

HOSPs have also been suggested in quasi-2D kagome superconductors $A$V$_3$Sb$_5$ \cite{hu2022rich,kang2022twofold}.
Coefficients of the quartic momentum-dependent term in their energy dispersions are significant ($\sim$13 eV$\cdot$\AA$^4$), resulting in a relatively dispersive band. To this perspective, the HOSP $\zeta_1$ of Co-based kagome materials is unique because of its relatively large area of flat band rooted in the small coefficient of the dominant quartic term, which is absent in other materials.

\textit{Conclusion.}
Our study reveals that HOSPs in 3D momentum space elevate the density of states by flattening the energy dispersion. We've identified two such 3D HOSPs in kagome metals YbCo$_6$Ge$_6$ and MgCo$_6$Ge$_6$, noted for their distinctive response to element substitution and lattice distortion. The unique presence of HOSPs, the induced flat band, and the Dirac point near the Fermi energy make these materials ideal candidates for probing correlation effects, such as electronic nematicity and Pomeranchuk instability.

\textit{Acknowledgements.}
B.Y. acknowledges the financial support from the European Research Council (ERC Consolidator Grant ``NonlinearTopo'', No. 815869), the ISF - Personal Research Grant	(No. 2932/21), and the DFG (CRC 183, A02).
J.Y.C. acknowledges Welch Foundation AA-2056-20220101 for partial support of this work.

%

\end{document}


\title{Supplemental Material for ``Three-dimensional higher-order saddle points induced flat bands in Co-based kagome metals''}
\author{Hengxin Tan}
\affiliation{Department of Condensed Matter Physics, Weizmann Institute of Science, Rehovot 7610001, Israel}

\author{Yiyang Jiang}
\affiliation{Department of Condensed Matter Physics, Weizmann Institute of Science, Rehovot 7610001, Israel}

\author{Gregory T. McCandless}
\author{Julia Y. Chan}
\affiliation{Department of Chemistry and Biochemistry, Baylor University, Waco, TX 76798, USA}
\author{Binghai Yan}
\affiliation{Department of Condensed Matter Physics, Weizmann Institute of Science, Rehovot 7610001, Israel}

\maketitle

This supplementary bears the following contents,
\begin{itemize}

    \item {Section \ref{SecI} shows the generation of higher-order saddle points (HOSPs) in the 2D kagome band structure involving the next-nearest neighboring hopping.}
    
    \item {Section \ref{SecII} presents a detailed derivation on the density of state of the 3D HOSPs and the scaling law of their first derivative.}
    
    \item {Section \ref{SecIII} shows the DFT calculation details and additional results not shown in the main text.}
    
\end{itemize}

\newpage
\section{2D Kagome model involving the next-nearest neighboring hoppings} \label{SecI}

\begin{figure}[tbp]
\includegraphics[width=0.9\linewidth]{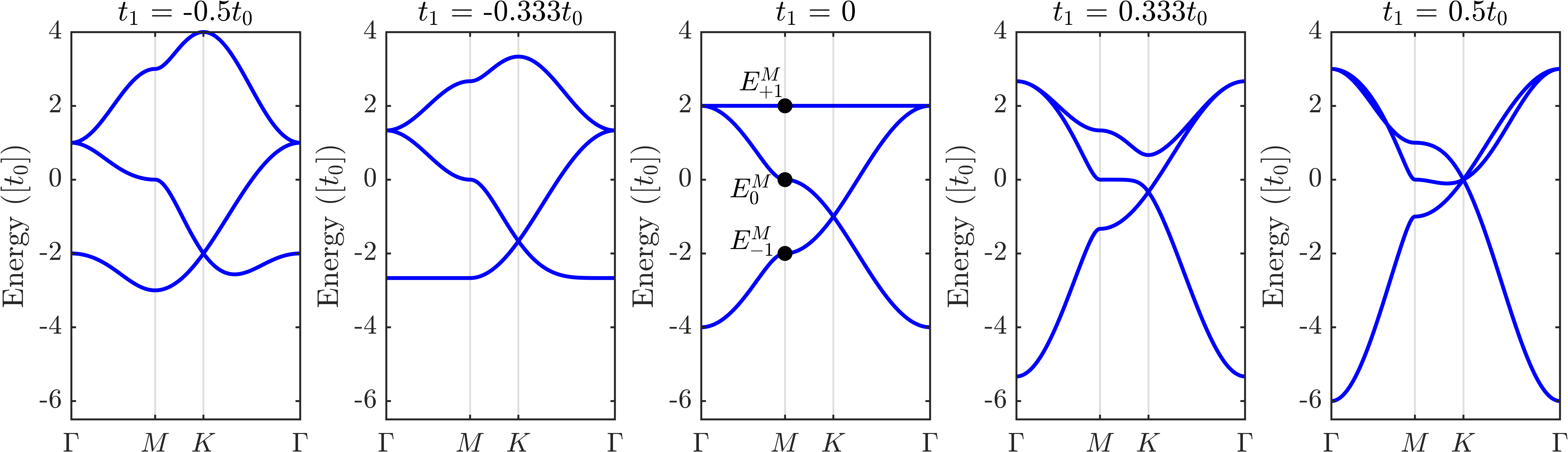}
\caption{ \label{FigS1} Evolution of the kagome model band structure with varying $t_1/t_0$ ratio. The $E_{0,\pm 1}^{M}$ notations used in the text are labeled.}
\end{figure}

In this section, we show the appearance of higher-order saddle points (HOSPs) in the 2D kagome model when the next-nearest neighboring hopping is considered. 
The Bloch Hamiltonian $H_0(\mathbf{k})$ of the kagome tight-binding model with nearest and next-nearest neighbor hoppings can be expressed as,
\begin{equation}
H_0(\mathbf{k})= 2 t_0 \left(
\begin{array}{ccc}
 0 & \cos \left(\mathbf{k}\cdot \mathbf{r}_1\right) & \cos \left(\mathbf{k}\cdot \mathbf{r}_3\right) \\
 \cos \left(\mathbf{k}\cdot \mathbf{r}_1\right) & 0 & \cos \left(\mathbf{k}\cdot \mathbf{r}_2\right) \\
 \cos \left(\mathbf{k}\cdot \mathbf{r}_3\right) & \cos \left(\mathbf{k}\cdot \mathbf{r}_2\right) & 0 \\
\end{array}
\right)+2 t_1 \left(
\begin{array}{ccc}
 0 & \cos \left(\mathbf{k}\cdot \mathbf{r}_{12}\right) & \cos \left(\mathbf{k}\cdot \mathbf{r}_{11}\right) \\
 \cos \left(\mathbf{k}\cdot \mathbf{r}_{12}\right) & 0 & \cos \left(\mathbf{k}\cdot \mathbf{r}_{13}\right) \\
 \cos \left(\mathbf{k}\cdot \mathbf{r}_{11}\right) & \cos \left(\mathbf{k}\cdot \mathbf{r}_{13}\right) & 0 \\
\end{array}
\right),
\label{eqS1}
\end{equation}
where $\mathbf{r}_1=a \left(\frac{1}{4},\frac{\sqrt{3}}{4}\right)$; $\mathbf{r}_2=a \left(\frac{1}{4},-\frac{\sqrt{3}}{4}\right)$; $\mathbf{r}_3=a \left(-\frac{1}{2},0\right)$ are the three vectors connecting the three nearest neighboring kagome atoms. $a$ is the lattice constant. The nearest-neighbor hopping parameter is $t_0$. Similarly, $\mathbf{r}_{11}=a \left(0,\frac{\sqrt{3}}{2}\right)$; $\mathbf{r}_{12}=a \left(\frac{3}{4},-\frac{\sqrt{3}}{4}\right)$; $\mathbf{r}_{13}=a \left(-\frac{3}{4},-\frac{\sqrt{3}}{4}\right)$ refer to next nearest neighbor hopping vectors with overlay integral $t_1$.

By numerics, we can get the band structure of this kagome model with different $t_1/t_0$ ratios. Some typical band structures are shown in Fig. \ref{FigS1}. In the following, we will detail the energy dispersions at M point using perturbation theory.

We consider $H_0(\mathbf{k})$ around M$ = -\frac{\pi}{a} \left(1,\frac{1}{\sqrt{3}}\right)$, and replace $\mathbf{k}$ in Eq. \eqref{eqS1} with $\mathbf{k} + $M, we get the following Bloch Hamiltonian,
\begin{equation}
\scriptsize
H(\mathbf{k}) = \left(
\begin{array}{ccc}
 0 & 2 \left(t_0 \sin \left(\frac{1}{4} a \left(k_x+\sqrt{3} k_y\right)\right)+t_1 \sin \left(\frac{1}{4} a \left(3 k_x-\sqrt{3} k_y\right)\right)\right) & 2 \left(t_0 \sin \left(\frac{a k_x}{2}\right)+t_1 \sin \left(\frac{1}{2} \sqrt{3} a k_y\right)\right) \\
 2 \left(t_0 \sin \left(\frac{1}{4} a \left(k_x+\sqrt{3} k_y\right)\right)+t_1 \sin \left(\frac{1}{4} a \left(3 k_x-\sqrt{3} k_y\right)\right)\right) & 0 & 2 t_0 \cos \left(\frac{1}{4} a \left(k_x-\sqrt{3} k_y\right)\right)-2 t_1 \cos \left(\frac{1}{4} a \left(3 k_x+\sqrt{3} k_y\right)\right) \\
 2 \left(t_0 \sin \left(\frac{a k_x}{2}\right)+t_1 \sin \left(\frac{1}{2} \sqrt{3} a k_y\right)\right) & 2 t_0 \cos \left(\frac{1}{4} a \left(k_x-\sqrt{3} k_y\right)\right)-2 t_1 \cos \left(\frac{1}{4} a \left(3 k_x+\sqrt{3} k_y\right)\right) & 0 \\
\end{array}
\right)
\end{equation}

We expand the Hamiltonian to 2nd order of $|\mathbf{k}|$ for perturbative analysis.
\begin{equation}
\begin{aligned}
H(\mathbf{k}) & \approx \left(
\begin{array}{ccc}
 0 & 0 & 0 \\
 0 & 0 & 2 \left(t_0-t_1\right) \\
 0 & 2 \left(t_0-t_1\right) & 0 \\
\end{array}
\right) \\
& + \left(
\begin{array}{ccc}
 0 & \frac{1}{2} a \left(\left(t_0+3 t_1\right) k_x+\sqrt{3} \left(t_0-t_1\right) k_y\right) & a \left(t_0 k_x+\sqrt{3} t_1 k_y\right) \\
 \frac{1}{2} a \left(\left(t_0+3 t_1\right) k_x+\sqrt{3} \left(t_0-t_1\right) k_y\right) & 0 & 0 \\
 a \left(t_0 k_x+\sqrt{3} t_1 k_y\right) & 0 & 0 \\
\end{array}
\right) \\
& + \left(
\begin{array}{ccc}
 0 & 0 & 0 \\
 0 & 0 & \frac{1}{16} a^2 \left(2 \sqrt{3} \left(t_0+3 t_1\right) k_x k_y-\left(t_0-9 t_1\right) k_x^2+3 \left(t_1-t_0\right) k_y^2\right) \\
 0 & \frac{1}{16} a^2 \left(2 \sqrt{3} \left(t_0+3 t_1\right) k_x k_y-\left(t_0-9 t_1\right) k_x^2+3 \left(t_1-t_0\right) k_y^2\right) & 0 \\
\end{array}
\right) \\
& + \mathcal{O}(|\mathbf{k}|^3) \\
& \equiv H_0 + H'
\end{aligned}
\end{equation}

Here, we define unperturbed Hamiltonian $H_0$ as the Bloch Hamiltonian at M point, and other terms are included in $H' = \mathcal{O}(|\mathbf{k}|)$. For $H_0$, the eigenvalues are,
\begin{equation}
E_0^{M} = 0,\, E_{\pm1}^{M} = \pm 2(t_0 - t_1),\\
\end{equation}
as shown in Fig. \ref{FigS1}.
If $t_1=0$, $E_0$ and $E_{-1}$ are the known pure- and mixed-type saddle points (SPs) and $E_{+1}$ corresponds to the flat band in the general kagome band structure. Their corresponding eigenvectors are,
\begin{equation}
| 0^{M} \rangle  = (1,0,0)^T,\, | \pm 1^{M} \rangle = \frac{{1}}{\sqrt{2}}(0,1,\pm1)^T.\\
\end{equation}
We focus on the SP $E^M_0$ in this work and further discussions on others will be presented in our independent work.

\subsection{$E_0^{M}$: 2nd order perturbation theory}\label{near-fermi-level}

It's easy to see the 1st order perturbation to $E^M_0$ is zero, so we consider 2nd order perturbation. Calculation of the perturbation matrix element gives,
\begin{equation}
\begin{aligned}
\langle 0^{M} | H' | +1^{M} \rangle &= \frac{a \left(t_0+t_1\right) \left(3 k_x+\sqrt{3} k_y\right)}{2 \sqrt{2}}\\
\langle 0^{M} | H' | -1^{M} \rangle &= -\frac{a \left(t_0-3 t_1\right) \left(k_x-\sqrt{3} k_y\right)}{2 \sqrt{2}}
\end{aligned}
\end{equation}

Then, 2nd order eigenvalue perturbation gives,
\begin{equation}
\begin{aligned}
E_0^{(M,2)} & = \sum_{m=\pm1}\frac{\langle 0^{M} | H' | m^{M} \rangle \langle m^{M} | H' | 0^{M} \rangle}{E_0^{M} - E_m^{M}} \\
& = \frac{|\langle 0^{M} | H' | +1^{M} \rangle|^2}{E_0^{M}-E_{+1}^{M}} + \frac{|\langle 0 | H' | -1 \rangle|^2}{E_0^{M} -E_{-1}^{M} } \\
& = -\frac{a^2 \left(t_0+t_1\right){}^2 \left(3 k_x+\sqrt{3} k_y\right){}^2}{16 \left(t_0-t_1\right)} + \frac{a^2 \left(t_0-3 t_1\right){}^2 \left(k_x-\sqrt{3} k_y\right){}^2}{16 \left(t_0-t_1\right)} \\
& = -\frac{a^2 \left(\sqrt{3} \left(t_0^2+3 t_1^2\right) k_x k_y+t_0 \left(t_0+3 t_1\right) k_x^2+3 \left(t_0-t_1\right) t_1 k_y^2\right)}{2 \left(t_0-t_1\right)}
\end{aligned}
\end{equation}

Rotating the coordinate clockwise by $\pi/3$, which changes $k_x$ axis into M-K direction, and $k_y$ axis into $\Gamma$-M direction, we get,
\begin{equation}
E_0^{(M,2)} = a^2\frac{\left(t_0-3 t_1\right){}^2 k_{\text{KM}}^2-3 \left(t_0+t_1\right){}^2 k_{\Gamma M}^2}{4 \left(t_0-t_1\right)} + \mathcal{O}(|\mathbf{k}|^3)
\end{equation}

From the explicit expression, we can see that when the next nearest hopping is 1/3 of the nearest hopping, i.e.,$t_1 = t_0/3$, then the coefficients of $k_{\Gamma M}$ becomes zero, making band dispersion independent of $k$ along M-K up to $\mathcal{O}(|\mathbf{k}|^2)$ order, which would be,
\begin{equation}
\left.E_0^{M}\right|_{t_1 = t_0/3} = -2 a^2 t_0 k_{\Gamma M}^2 + \mathcal{O}(|\mathbf{k}|^3).
\end{equation}

\subsubsection{$E_0^{M}$: 4th order perturbation theory}\label{th-order-perturbation-theory-m}

We now consider perturbation theory higher than 2nd order to get more information about the SP. The general formula for 3rd-order and 4th-order perturbation to eigenvalue is,

\begin{equation}
\begin{aligned}
E_0^{(M,3)} = -\sum_{l=\pm1}  \frac{H'_{0l}H'_{l0}H'_{00}}{(E_0-E_l)^2} + \sum_{l=\pm1}\sum_{l'=\pm1}  \frac{H'_{0l}H'_{ll'}H'_{l'0}}{(E_0-E_l)(E_0-E_{l'})}
\end{aligned}
\end{equation}
and
\begin{equation}
\begin{aligned}
E_0^{(M,4)} & = \sum_{l=\pm1} \frac{H'_{0l}H'_{l0}H'_{00}H'_{00}}{(E_0-E_l)^3} \\
& - \sum_{l=\pm1}\sum_{l'=\pm1}  \frac{H'_{0l}H'_{ll'}H'_{l'0}H'_{00}}{(E_0-E_l)(E_0-E_{l'})} \left( \frac{1}{E_0-E_l} + \frac{1}{E_0-E_{l'}} \right) \\
& - \frac{5}{6} \sum_{l=\pm1}\sum_{l'=\pm1} \frac{H'_{0l}H'_{l0}H'_{0l'}H'_{l'0}}{(E_0-E_l)(E_0-E_{l'})} \left( \frac{1}{E_0-E_l} + \frac{1}{E_0-E_{l'}} \right) \\
& + \sum_{l=\pm1}\sum_{l'=\pm1}\sum_{l''=\pm1} \frac{H'_{0l}H'_{ll'}H'_{l'l''}H'_{l''0}}{(E_0-E_l)(E_0-E_{l'})(E_0-E_{l''})}
\end{aligned}
\end{equation}

where $H'_{lm} = \langle l | H' | m \rangle$. We need to expand $H'$ to $\mathcal{O}(|\mathbf{k}|^4)$,
\begin{equation}
\begin{aligned}
H' & = \left(
\begin{array}{ccc}
 0 & \frac{1}{2} a \left(\sqrt{3} \left(t_0+t_1\right) k_{\Gamma M}-\left(t_0-3 t_1\right) k_{\text{KM}}\right) & \frac{1}{2} a \left(\left(t_0-3 t_1\right) k_{\text{KM}}+\sqrt{3} \left(t_0+t_1\right) k_{\Gamma M}\right) \\
 \frac{1}{2} a \left(\sqrt{3} \left(t_0+t_1\right) k_{\Gamma M}-\left(t_0-3 t_1\right) k_{\text{KM}}\right) & 0 & 0 \\
 \frac{1}{2} a \left(\left(t_0-3 t_1\right) k_{\text{KM}}+\sqrt{3} \left(t_0+t_1\right) k_{\Gamma M}\right) & 0 & 0 \\
\end{array}
\right) \\
& + \left(
\begin{array}{ccc}
 0 & 0 & 0 \\
 0 & 0 & -\frac{1}{4} a^2 \left(t_0 k_{\text{KM}}^2-3 t_1 k_{\Gamma M}^2\right) \\
 0 & -\frac{1}{4} a^2 \left(t_0 k_{\text{KM}}^2-3 t_1 k_{\Gamma M}^2\right) & 0 \\
\end{array}
\right) \\
& + \left(
\begin{array}{ccc}
 0 & \frac{1}{192} a^3 \left\{ \begin{array}{c} -3 \sqrt{3} \left(t_0+9 t_1\right) k_{\Gamma M} k_{\text{KM}}^2 \\ +9 \left(t_0-3 t_1\right) k_{\Gamma M}^2 k_{\text{KM}} \\ +\left(t_0-27 t_1\right) k_{\text{KM}}^3 \\ -3 \sqrt{3} \left(t_0+t_1\right) k_{\Gamma M}^3 \end{array} \right\} & -\frac{1}{192} a^3 \left\{  \begin{array}{c} 3 \sqrt{3} \left(t_0+9 t_1\right) k_{\Gamma M} k_{\text{KM}}^2\\ +9 \left(t_0-3 t_1\right) k_{\Gamma M}^2 k_{\text{KM}}\\ +\left(t_0-27 t_1\right) k_{\text{KM}}^3\\ +3 \sqrt{3} \left(t_0+t_1\right) k_{\Gamma M}^3 \end{array} \right\} \\
 \frac{1}{192} a^3 \left\{ \begin{array}{c} -3 \sqrt{3} \left(t_0+9 t_1\right) k_{\Gamma M} k_{\text{KM}}^2 \\ +9 \left(t_0-3 t_1\right) k_{\Gamma M}^2 k_{\text{KM}} \\ +\left(t_0-27 t_1\right) k_{\text{KM}}^3\\ -3 \sqrt{3} \left(t_0+t_1\right) k_{\Gamma M}^3 \end{array} \right\} & 0 & 0 \\
 -\frac{1}{192} a^3 \left\{  \begin{array}{c} 3 \sqrt{3} \left(t_0+9 t_1\right) k_{\Gamma M} k_{\text{KM}}^2\\ +9 \left(t_0-3 t_1\right) k_{\Gamma M}^2 k_{\text{KM}}\\ +\left(t_0-27 t_1\right) k_{\text{KM}}^3\\ +3 \sqrt{3} \left(t_0+t_1\right) k_{\Gamma M}^3 \end{array} \right\} & 0 & 0 \\
\end{array}
\right)\\
& + \left(
\begin{array}{ccc}
 0 & 0 & 0 \\
 0 & 0 & \frac{1}{192} a^4 \left(t_0 k_{\text{KM}}^4-9 t_1 k_{\Gamma M}^4\right) \\
 0 & \frac{1}{192} a^4 \left(t_0 k_{\text{KM}}^4-9 t_1 k_{\Gamma M}^4\right) & 0 \\
\end{array}
\right) \\
& + \mathcal{O}(|\mathbf{k}|^5)
\end{aligned}
\end{equation}

After tedious calculation, we find,
\begin{equation}
\begin{aligned}
E_0^{M} & = E_0^{(M,0)} + E_0^{(M,1)} + E_0^{(M,2)} + E_0^{(M,3)} + E_0^{(M,4)} + \mathcal{O}(|\mathbf{k}|^5) \\
& = \frac{a^2 \left(\left(t_0-3 t_1\right){}^2 k_{\text{KM}}^2-3 \left(t_0+t_1\right){}^2 k_{\Gamma M}^2\right)}{4 \left(t_0-t_1\right)} \\
& -\frac{a^4 \left(18 \left(t_0-3 t_1\right){}^2 \left(t_0^2-t_1^2\right) k_{\Gamma M}^2 k_{\text{KM}}^2+\left(t_0^4-62 t_1 t_0^3+376 t_1^2 t_0^2-786 t_1^3 t_0+567 t_1^4\right) k_{\text{KM}}^4-9 \left(t_0+t_1\right){}^2 \left(7 t_0^2+16 t_1 t_0+t_1^2\right) k_{\Gamma M}^4\right)}{192 \left(t_0-t_1\right){}^3} \\
& + \mathcal{O}(|\mathbf{k}|^6) \\
& = \frac{a^2 \left(\left(t_0-3 t_1\right){}^2 k_{\text{KM}}^2-3 \left(t_0+t_1\right){}^2 k_{\Gamma M}^2\right)}{4 \left(t_0-t_1\right)} \\
& -\frac{a^4 \left(18 \left(t_0-3 t_1\right){}^2 \left(t_0^2-t_1^2\right) k_{\Gamma M}^2 k_{\text{KM}}^2+ \left(t_0-3 t_1\right) \left(t_0^3-59 t_1 t_0^2+199 t_1^2 t_0-189 t_1^3\right) k_{\text{KM}}^4-9 \left(t_0+t_1\right){}^2 \left(7 t_0^2+16 t_1 t_0+t_1^2\right) k_{\Gamma M}^4\right)}{192 \left(t_0-t_1\right){}^3} \\
& + \mathcal{O}(|\mathbf{k}|^6)
\end{aligned}
\end{equation}

The 5th order of $k$ would also vanish due to symmetry constraints, just like the 1st and 3rd orders. Thus, the result is accurate up to the 6th order of $k$. We find that when $t_1 = \frac{1}{3}t_0$, not only the coefficient of $k_{\text{KM}}^2$ vanishes, but coefficient of $k_{\text{KM}}^4$
vanishes as well, i.e.
\begin{equation}
\left.E_0^{M}\right|_{t_1 = t_0/3} = -2 a^2 t_0 k_{\Gamma M}^2 - \frac{7}{2} a^4 t_0 k_{\Gamma M}^4 + \mathcal{O}(|\mathbf{k}|^6).
\end{equation}
This expression corresponds to a HOSP because the lowest order of $k_{MK}$ will be at least 6, which justifies the flatness of the band along the M-K line.

\newpage

\begin{figure}[tbp]
\includegraphics[width=0.95\linewidth]{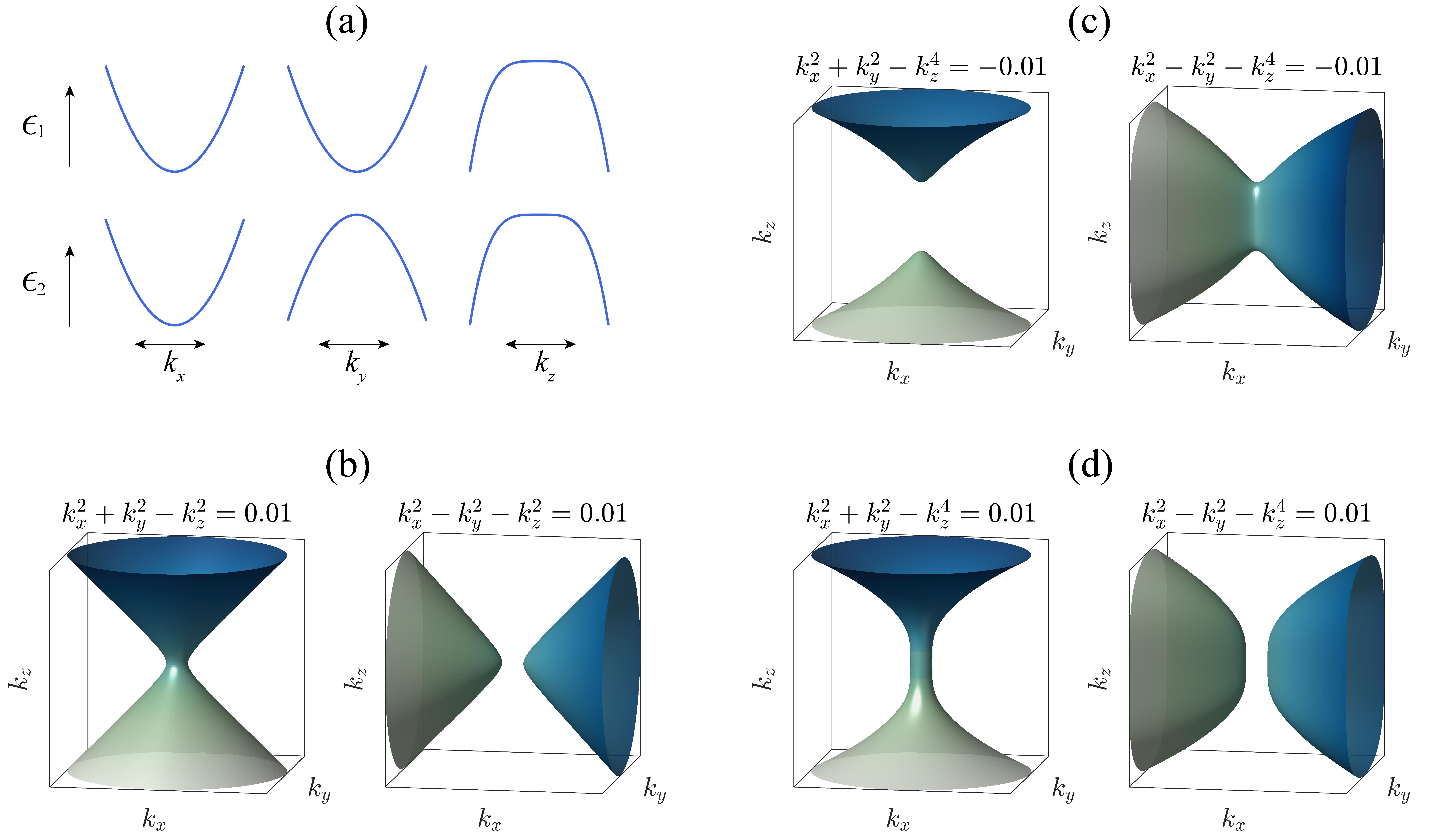}
\caption{ \label{FigSII} (a) shows the energy dispersion along three different directions of two HOSPs $\epsilon_1$ and $\epsilon_2$. For plotting, we have set $n=4$. (b) depicts the constant energy contour ($\epsilon=0.01$) of the general saddle point ($n=2$). (c) and (d) presents the constant energy contour plots of the two HOSPs ($n=4$) at $\pm$0.01, respectively. In all plots, we have set $\epsilon_0 = 0$.
}
\end{figure}

\section{DOS of the HOSP} \label{SecII}

The simplest HOSP can take the following two forms, as we show in the main text,
\begin{equation}
\begin{cases}
 \epsilon_1(\textbf{k}) =& \epsilon_0 + c_x k_x^2 + c_y k_y^2 - c_{z} k_z^{n}, \\
 \epsilon_2(\textbf{k}) =& \epsilon_0 +  c_x k_x^2 - c_y k_y^2 + c_{z} k_z^{n}. \\
\end{cases}
\label{eq2}
\end{equation}
We have set the coefficients {$c_x$, $c_y$, $c_z$} to positive values. We constrain $n$ to only even integers in consideration of the mirror symmetries of the Brillouin zone boundary of the crystals. $n=2$ corresponds to the ordinary saddle points.

Figure \ref{FigSII}(a) shows the energy dispersion of two HOSPs along three perpendicular directions. The higher-order momentum dependence gives rise to a flatter dispersion near the saddle point, as indicated by the dispersion along the $k_z$ direction. Fig. \ref{FigSII}(c)\&(d) depict the constant energy contours of two HOSPs ($n=4$) at $\pm$0.01, which shows a clear difference with that of the ordinary SPs in Fig. \ref{FigSII}(b). We have set $\epsilon_0 = 0$ in these plots. The Fermi surfaces, i.e., constant energy contour at $\epsilon_0$ (zero energy) are presented in the main text.

In the following, we derive the density of states (DOS) and their first-derivative for the two HOSPs utilizing the Dirac delta function: $g(E) = \frac{1}{(2\pi)^3} \int \delta(E-\epsilon) d\textbf{k}$.

\subsection{ DOS of $\epsilon_1 = \epsilon_0 + c_x k_x^2 + c_y k_y^2 - c_{z} k_z^{n}$}

For $\epsilon_1$, re-write $k_x$ and $k_y$ in terms of polar coordinates, we have $k_x = \frac{q}{\sqrt c_x}$ cos($\theta$) and $k_y = \frac{q}{\sqrt c_y}$ sin($\theta$), where $q$ and $\theta$ are the radial coordinate and angular coordinate, respectively. We also re-scale $k_z$ to $\frac{1}{\sqrt[n]{c}}k_z$. Then we have $\epsilon_1 =  \epsilon_0 + q^2 - k_z^{n}$. The DOS is 
\begin{equation}
\begin{aligned}
 g_1(E) &= \frac{1}{(2\pi)^3} \int \delta(E-\epsilon_1) d\textbf{k} \\
      &= \frac{1}{(2\pi)^3} \int_q \int_{\theta} \int_{k_z} \delta(E-\epsilon_1) q dq \cdot d\theta \cdot \frac{1}{\sqrt[n]{c_z}}dk_z\\
      &= \frac{1}{(2\pi)^3} \frac{2\pi}{\sqrt{c_xc_y} \sqrt[n]{c_z}}  \int_q \int_{k_z} \delta(E-\epsilon_0 -q^2 + k_z^n) q dq \cdot dk_z\\
      &= \frac{1}{(2\pi)^3} \frac{2\pi}{\sqrt{c_xc_y} \sqrt[n]{c_z}}  \int_q  dq \sum_{k_z} \left. \frac{q}{|nk_z^{n-1}|} \right|_{E-\epsilon_0-q^2+k_z^n = 0}\\
      &= \frac{1}{(2\pi)^3} \frac{2\pi}{\sqrt{c_xc_y} \sqrt[n]{c_z}}  \int_q  dq \frac{2q}{n (\epsilon_0+q^2-E)^{\frac{n-1}{n}}} \\
      &= \frac{1}{(2\pi)^3} \frac{2\pi}{\sqrt{c_xc_y}} \frac{1}{n \sqrt[n]{c_z}}  \int_q (\epsilon_0+q^2-E)^{\frac{1}{n}-1} d(q^2)\\
      &= \frac{1}{4\pi^2\sqrt{c_xc_y}} \frac{1}{n \sqrt[n]{c_z}}  \left. [n (\epsilon_0+q^2-E)^{\frac{1}{n}}]\right|_{q_l}^{R} \\
\end{aligned}
\end{equation}
Notice that the $\sum_{k_z}$ runs over all $k_z$ which satisfy $E-\epsilon_0-q^2+k_z^n = 0$. $R$ defines the radius of the upper limit of $q$ (we are only interested in DOS from the HOSP). $q_l$ defines the lower boundary of the integral range for $q$ ($\geq0$). Depending on $E$, $q_l$ has different values.

\noindent \textbf{i) $E \leq \epsilon_0$: $q_l = 0$}.
\begin{equation}
\begin{aligned}
 g_1(E) &= \frac{1}{4\pi^2 \sqrt{c_xc_y}} \frac{1}{\sqrt[n]{c_z}}  \left( \sqrt[n]{\epsilon_0+R^2-E} - \sqrt[n]{\epsilon_0-E} \right)\\
\end{aligned}
\end{equation}

\noindent \textbf{ii) $\epsilon_0 < E \leq \epsilon_0+R^2$: $q_l= \sqrt{E-\epsilon_0}$.} The lower boundary $q_l = \sqrt{E-\epsilon_0}$ comes from the condition $k_z^n = \epsilon_0+q^2-E \geq 0$ because $n$ is even.
\begin{equation}
\begin{aligned}
 g_1(E) &= \frac{1}{4\pi^2 \sqrt{c_xc_y}} \frac{1}{\sqrt[n]{c_z}}  \sqrt[n]{\epsilon_0+R^2-E} \\
\end{aligned}
\end{equation}

\noindent \textbf{iii) $E > \epsilon_0+R^2$:} $k_z^n = \epsilon_0+q^2-E \geq 0$ has no solution because $q \leq R$. Thus the DOS is zero.

The 1st-order derivative at $\epsilon_0$ is
\begin{equation}
\begin{aligned}
 g_1'(E) = \frac{1}{4\pi^2 \sqrt{c_xc_y}} \frac{1}{\sqrt[n]{c_z}} 
 \begin{cases}
 \frac{1}{n} \left[ -R^{\frac{2}{n}-2} +  \left( \epsilon_0-E\right)^{\frac{1}{n}-1} \right], &(E \rightarrow \epsilon_0^-)\\
 -\frac{1}{n} R^{\frac{2}{n}-2} , &(E \rightarrow \epsilon_0^+)\\
 \end{cases}
\end{aligned}
\label{g1-1st}
\end{equation}

\subsection{ DOS of $\epsilon_2(\textbf{k}) = \epsilon_0 +  c_x k_x^2 - c_y k_y^2 - c_{z} k_z^{n}$ }

For $\epsilon_2$, we do a similar transformation, $k_y = \frac{q}{\sqrt c_y}$ cos($\theta$) and $k_z = \pm \frac{q^{\frac{2}{n}}}{\sqrt[n]{c_z}}$ sin$^{\frac{2}{n}}$($\theta$), where $q$ and $\theta$ ($\in [0, \pi)$) are the radial and angular coordinate, respectively. We re-scale $k_x$ to $\frac{1}{\sqrt{c_x}}k_x$ as well. Then we have $\epsilon_2 =  \epsilon_0 + k_x^2 - q^2$.

Since $\epsilon_2$ is an even function with respect to $k_z$, the DOS contribution from $k_z<0$ is the same as that from $k_z>0$. The total DOS will be obtained by multiplying a factor 2 to the DOS from $k_z>0$ part, i.e., $k_z = \frac{q^{\frac{2}{n}}}{\sqrt[n]{c_z}}$ sin$^{\frac{2}{n}}$($\theta$), which we will derive below. The length differentials along $k_y$ and $k_z$ are
\begin{equation}
 \begin{cases}
 d \vec{k_y} = \frac{1}{\sqrt{c_y}} \left[ \text{cos}(\theta) d\vec{q} - q \text{sin}(\theta) d \vec{\theta} \right] \\
 d \vec{k_z} = \frac{1}{\sqrt[n]{c_z}} \frac{2}{n} \left[ q^{\frac{2}{n}-1}\text{sin}^{\frac{2}{n}}(\theta) d\vec{q} + q^{\frac{2}{n}} \text{sin}^{\frac{2}{n}-1}(\theta) \text{cos}(\theta) d \vec{\theta} \right] \\
 \end{cases}
\end{equation}

\noindent The area differential of $|d \vec{k_y} \times d \vec{k_z}|$ can be expressed as 
\begin{equation}
\begin{aligned}
|d \vec{k_y} \times d \vec{k_z}| &= \frac{1}{\sqrt{c_x} \sqrt[n]{c_z}} \frac{2}{n} \left[ q^{\frac{2}{n}} \text{sin}^{\frac{2}{n}-1}(\theta) \text{cos}^2(\theta) +  q^{\frac{2}{n}}\text{sin}^{\frac{2}{n}+1}(\theta) \right] dq d \theta \\
 &= \frac{1}{\sqrt{c_y} \sqrt[n]{c_z}} \frac{2}{n} q^{\frac{2}{n}} \text{sin}^{\frac{2}{n}-1}(\theta) dq d \theta\ .\
\end{aligned}
\end{equation}

The DOS is expressed as 
\begin{equation}
\begin{aligned}
 g_2(E) &= 2\times \frac{1}{(2\pi)^3} \int \delta(E-\epsilon_2) d\textbf{k} \\
        &= 2\times \frac{1}{(2\pi)^3} \int_q \int_{\theta} \int_{k_x} \delta(E-\epsilon_2) \cdot \frac{1}{\sqrt{c_y} \sqrt[n]{c_z}} \frac{2}{n} q^{\frac{2}{n}} \text{sin}^{\frac{2}{n}-1}(\theta) dq d \theta \cdot \frac{1}{\sqrt{c_x}}dk_x\\
        &= 2\times \frac{1}{(2\pi)^3} \frac{1}{\sqrt{c_xc_y} \sqrt[n]{c_z}} \frac{2}{n}  \int_{0}^{\pi}\text{sin}^{\frac{2}{n}-1}(\theta) d\theta  \cdot \int_q dq \left[q^{\frac{2}{n}} \int_{k_x} \delta(E-\epsilon_0 + q^2 - k_x^2) dk_x \right]\\
        &= \frac{1}{(2\pi)^3} \frac{1}{\sqrt{c_xc_y} \sqrt[n]{c_z}} \frac{4}{n}  \int_{0}^{\pi}\text{sin}^{\frac{2}{n}-1}(\theta) d\theta  \cdot \int_q dq \left[q^{\frac{2}{n}} \Theta(q^2 + E - \epsilon_0) \sum_{k_x} \left. \frac{1}{2|k_x|} \right|_{E-\epsilon_0 + q^2 - k_x^2 = 0} \right]\\
        &= \frac{1}{4\pi^2} \frac{1}{\sqrt{c_xc_y} \sqrt[n]{c_z}} \cdot \frac{1}{2\pi} \frac{4}{n}  \int_{0}^{\pi}\text{sin}^{\frac{2}{n}-1}(\theta) d\theta  \cdot \int_{q} dq \frac{q^{\frac{2}{n}}}{\sqrt{q^2+E-\epsilon_0}} \Theta(q^2+E-\epsilon_0),
\end{aligned}
\end{equation}
where $\Theta(x)$ is the Heaviside step function. The integral of $\theta$ can be carried out as,
\begin{equation}
\begin{aligned}
 \int_{0}^{\pi}\text{sin}^{\frac{2}{n}-1}(\theta) d\theta &= 2 \int_{0}^{\pi/2}\text{sin}^{\frac{2}{n}-1}(\theta) d\theta \\
 &\xlongequal[d\theta = \frac{dt}{\sqrt{1-t^2}}]{t = \sin \theta} 2 \int_0^1 t^{\frac{2}{n}-1} (1 - t^2)^{-\frac{1}{2}} d t \\
 &\xlongequal[dt = \frac{du}{2 \sqrt{u}}]{u = t^2} 2 \int_0^1 (\sqrt{u})^{\frac{2}{n}-1} (1 - u)^{-\frac{1}{2}} \frac{du}{2 \sqrt{u}} \\
 & = \int_0^1 u^{\frac{1}{n} - 1} (1-u)^{-\frac{1}{2}} du \\
 & = B(\frac{1}{n}, \frac{1}{2}) \\
 & = \frac{ \Gamma( \frac{1}{n} ) \Gamma( \frac{1}{2} ) }{\Gamma( \frac{1}{n} + \frac{1}{2} )}
\end{aligned}
\end{equation}
where $\Gamma(x)$ is the Gamma function.

The integral of $q$ can be further derived,
\begin{equation}
\begin{aligned}
 G_n(E) \equiv \int_{0}^{R} dq \frac{q^{\frac{2}{n}}}{\sqrt{q^2+E-\epsilon_0}} \Theta(q^2+E-\epsilon_0) & = \left\{
 \begin{aligned}
 & \int_{0}^{R} dq \frac{q^{\frac{2}{n}}}{\sqrt{q^2+E-\epsilon_0}} ,\,\epsilon_0 -E < 0 \\
 & \int_{\sqrt{\epsilon_0-E}}^{R} dq \frac{q^{\frac{2}{n}}}{\sqrt{q^2+E-\epsilon_0}} ,\,0 < \epsilon_0 -E < R^2 \\
 & 0,\, R^2 < \epsilon_0 - E
\end{aligned}
 \right. \\
 & = \left\{
 \begin{aligned}
 & \int_{0}^{R} dq \frac{q^{\frac{2}{n}}}{\sqrt{q^2+E-\epsilon_0}} ,\, \epsilon_0 -E< 0 \\
 & \int_{0}^{\sqrt{R^2 - (\epsilon_0-E)}} dq' \left( {q'}^2 + \epsilon_0 -E\right)^{\frac{1}{n} - \frac{1}{2}} ,\,0< \epsilon_0 -E < R^2\\
 & 0,\, R^2 < \epsilon_0 - E
\end{aligned}
 \right.
\end{aligned}
\end{equation}

Consider the behavior of $G_n(E)$ when $E \rightarrow \epsilon_0$
\begin{equation}
 \lim_{E\rightarrow \epsilon_0^+} G_n(E) = \lim_{E\rightarrow \epsilon_0^-} G_n(E) = \int_0^{R} q^{\frac{2}{n} - 1} dq = \frac{n}{2} R^{\frac{2}{n}}.
\end{equation}
This means that $G_n(E)$ is a finite value. Considering the integral of $\theta$ is also finite, $g_2(E)$ is finite.

The 1st-order derivative of $G_n(E)$ at $\epsilon_0$ reads, 
\begin{equation}
\begin{aligned}
 \lim_{E\rightarrow \epsilon_0^+} \frac{\partial}{\partial E} G_n(E) & = \int_0^{R} dq q^{\frac{2}{n}} \left. \frac{\partial}{\partial E} (q^2 +E - \epsilon_0)^{-\frac{1}{2}} \right|_{E = \epsilon_0^+} \\
 & = -\frac{1}{2} \int_0^{R} dq q^{\frac{2}{n} - 3} \\
 & = -\frac{1}{2} \frac{1}{\frac{2}{n} - 2} \left. q^{\frac{2}{n} - 2} \right|_0^{R} \\
 & = \frac{1}{4} \frac{n}{n-1} \left. q^{\frac{2}{n} - 2} \right|^{R}_0 \\
 & = - \infty,\, (n>1)
\\
\\
 \lim_{E\rightarrow \epsilon_0^-} \frac{\partial}{\partial E} G_n(E) & = \int_0^{\sqrt{R^2 - (\epsilon_0-E)}} dq' \left. \frac{\partial}{\partial E}  \left( {q'}^2 + \epsilon_0 - E \right)^{\frac{1}{n} - \frac{1}{2}} \right|_{E = \epsilon_0^-} \\
 & + \left. \frac{\partial}{\partial E} \left( \sqrt{R^2 - (\epsilon_0-E)} \right) \right|_{E = \epsilon_0^-} \cdot \left. \left( {q'}^2 +\epsilon_0-E \right)^{\frac{1}{n} - \frac{1}{2}} \right|_{q' =\sqrt{R^2 - (\epsilon_0-E)},\, E = \epsilon_0^-} \\
 & = - (\frac{1}{n} - \frac{1}{2}) \int_0^{R} dq' \left( q' \right)^{\frac{2}{n} - 3} + \frac{1}{2 R} \cdot R^{\frac{2}{n}-1} \\
 & = \frac{n-2}{4(n-1)} \left. q^{\frac{2}{n} - 2} \right|_0^{R} + \frac{1}{2} R^{\frac{2}{n}-2} \\
 & = \left\{
 \begin{aligned}
    & \frac{1}{2R},\, n=2 \\
    & +\infty ,\, n>2
 \end{aligned}
 \right.
\end{aligned}
\end{equation}

Notice that, when $n=2$, the integral of $\theta$ is $\pi$; the 1st-derivative of $g_2(E)$ at $\epsilon_0$ is 
\begin{equation}
\begin{aligned}
 g_2'(E)=\frac{1}{4\pi^2} \frac{1}{\sqrt{c_xc_y} \sqrt[n]{c_z}} 
 \begin{cases}
   \frac{1}{2} \left( \frac{1}{R} - \left. \frac{1}{q}\right|_{q\rightarrow0^+} \right), & (E\rightarrow\epsilon_0^+)\\
     \frac{1}{2R}, & (E\rightarrow\epsilon_0^-)
 \end{cases}
\end{aligned}
\end{equation}
in line with the 1st-order derivative of $g_1(E)$ in Eq. \eqref{g1-1st}, considering that $\epsilon_2-\epsilon_0 = -(\epsilon_1-\epsilon_0)$ when $n=2$.

\subsection{ Universal scaling law of HOSP DOS near $\epsilon_0$}
Now, we discuss the universal scaling law of the DOS near $\epsilon_0$.
In our case, energy dispersions of both $\epsilon_1$ and $\epsilon_2$ can be expressed as $\epsilon_i(k_x,k_y,k_z)-\epsilon_0 = c_x k_x^{n_x} + c_y k_y^{n_y} + c_z k_z^{n_z}$ that has scaling invariance:

\begin{equation}
    \epsilon_i(\lambda^{1/n_x}k_x, \lambda^{1/n_y}k_y, \lambda^{1/n_z} k_z ) - \epsilon_0 = \lambda \left[\epsilon_i(k_x, k_y, k_z ) - \epsilon_0\right], \quad \lambda > 1
\end{equation}

Then the scaling law of DOS can be derived as,

\begin{equation}
    \begin{aligned}
        g_i(\lambda (E-\epsilon_0) + \epsilon_0) = & \frac{1}{(2 \pi)^3} \int^R dk_x dk_y dk_z \delta( \lambda (E -\epsilon_0) + \epsilon_0 - \epsilon_i( k_x, k_y, k_z ) ) \\
        \xlongequal[]{\delta(\lambda x) = \frac{\delta(x)}{|\lambda|}} & \lambda^{-1} \frac{1}{(2 \pi)^3} \int^R dk_x dk_y dk_z \delta(  E -\epsilon_0 - \lambda^{-1} ( \epsilon_i( k_x, k_y, k_z )- \epsilon_0)  ) \\
        = & \lambda^{-1} \frac{1}{(2 \pi)^3} \int^R dk_x dk_y dk_z \delta(  E -\epsilon_0 - (\epsilon_i(\lambda^{-1/n_x} k_x, \lambda^{-1/n_y} k_y, \lambda^{-1/n_z} k_z )- \epsilon_0)  ) \\
        \xlongequal[R\rightarrow\infty]{k_{x,y,z}' = \lambda^{-1/n_{x,y,z}}k_{x,y,z}} & \lambda^{\frac{1}{n_x} + \frac{1}{n_y} + \frac{1}{n_z} - 1}  \frac{1}{(2 \pi)^3} \int^R dk_x' dk_y' dk_z' \delta(  E - \epsilon_i( k_x',  k_y',  k_z' ) ) \\
        = & \lambda^{\frac{1}{n_x} + \frac{1}{n_y} + \frac{1}{n_z} - 1} g_i((E -\epsilon_0) + \epsilon_0).
    \end{aligned}
\end{equation}

Therefore, at energy $\epsilon_0$, the DOS behaves as,

\begin{equation}
g_i(E) \propto |E - \epsilon_0|^{\frac{1}{n_x} + \frac{1}{n_y} + \frac{1}{n_z} - 1}
\end{equation}

In our case, $n_x = n_y = 2,\,n_z = n$, therefore the power-law scalings of DOS and its first derivative are,

\begin{equation}
g_i(E) \propto |E - \epsilon_0|^{\frac{1}{n}}, \quad g_i'(E) \propto |E - \epsilon_0|^{\frac{1}{n}-1}
\end{equation}

Since $n>1$, the DOS $g_i$ converges to a finite value at $\epsilon_0$ as the power $\frac{1}{n}>0$ while its first derivative is constantly diverging at $\epsilon_0$ as $\frac{1}{n}-1<0$.

\newpage
\section{DFT calculations} \label{SecIII}
\textit{Methodology.}
Electronic structures are calculated with the Density functional theory (DFT) implemented in the Vienna $ab$-$initio$ Simulation Package \cite{VASP}. The generalized gradient approximation parameterized by Perdew, Burke, and Ernzerhof (PBE) \cite{Perdew1997generalized} is utilized to mimic the exchange-correlation interaction between electrons throughout.
The projected augmented potential does not consider the $f$ electrons of rare-earth elements.
The energy cutoff for the plane wave basis set is 350 eV.
The crystal structures are all fully relaxed so that forces on atoms are less than 1 meV/\AA.
A $k$-mesh of 10$\times$10$\times$6 is used to sample the Brillouin zone.
Electronic structures are calculated with spin-orbital coupling.
The total density of states (DOS) is calculated using the tetrahedron method with Bl\"ochl corrections.
The DOS of the two higher-order saddle points are calculated with the wannier Hamiltonian, obtained from the wannier90 software \cite{wannier90}. The Co $d$ and Ge $p$ orbits are included in the wannier basis set. Phonon dispersions are calculated with phonopy software \cite{phonopy}. A supercell of 3$\times$3$\times$2 is employed.

Table \ref{tbA1} shows crystal structures of YbCo$_6$Ge$_6$ in different phases as obtained from DFT calculations.
More results not shown in the main text are shown below.

\begin{table}[h]
\centering
\caption{\label{Table_str} Lattice parameters ($a,b,c$) and Wyckoff positions of YbCo$_6$Ge$_6$ and MgCo$_6$Ge$_6$ in different phases. Since all phase are hexagonal, ($\alpha, \beta, \gamma$)=(90$^\circ$, 90$^\circ$, 120$^\circ$) are the same for all structures. SG = space group. The $P6_3/mcm$ (SG193) phase is the 1$\times$1$\times$2 CDW phase.
}
\renewcommand\arraystretch{1.3}
\begin{ruledtabular}
\begin{tabular}{c|l|l|l|l} \label{tbA1}
  & Pristine: $P6/mmm$ (SG191) & $P\bar{3}1m$ (SG162) & $P\bar{6}2m$ (SG189) & CDW: $P6_3/mcm$ (SG193) \\
  \hline
  \multirow{7}{*}{YbCo$_6$Ge$_6$}& $a = b = 5.087$ \AA,  $c = 7.827$ \AA  & $a=b=5.093$ \AA, $c = 7.818$ \AA  & $a=b=5.093$ \AA, $c = 7.818$ \AA  & $a=b=5.093$ \AA, $c = 15.635$ \AA \\
  &&&&\\
  &Yb1 1a (0, 0, 0)         & Yb1 1a (0, 0, 0)           & Yb1 1a (0, 0, 0)            & Yb1 2a (0, 0, 1/4)\\
  &Co1 6i (0.5, 0, 0.25098) & Co1 6k (0.47643, 0, 0.25073)& Co2 6i (0.47916, 0, 0.25088)& Co1 12k (0.47694, 0, 0.37534)\\
  &Ge1 2c (1/3, 2/3, 0)     & Ge1 2c (1/3, 2/3, 0)       & Ge1 2c (1/3, 2/3, 0)        & Ge1 4c (1/3, 2/3, 1/4)\\
  &Ge2 2e (0, 0, 0.34311)   & Ge2 2e (0, 0, 0.34320)     & Ge2 2e (0, 0, 0.34319)      & Ge2 4e (0, 0, 0.42160)\\
  &Ge3 2d (1/3, 2/3, 1/2)   & Ge3 2d (1/3, 2/3, 1/2)     & Ge3 2d (1/3, 2/3, 1/2)      & Ge3 4d (1/3, 2/3, 0)\\
  \hline
  \hline
  \multirow{7}{*}{MgCo$_6$Ge$_6$}& $a = b = 5.066$ \AA,  $c = 7.730$ \AA  & $a=b=5.072$ \AA, $c = 7.722$ \AA  & $a=b=5.076$ \AA, $c = 7.715$ \AA  & $a=b=5.076$ \AA, $c = 15.434$ \AA \\
  &&&&\\
  &Mg1 1a (0, 0, 0)         & Mg1 1a (0, 0, 0)           & Mg1 1a (0, 0, 0)            & Mg1 2a (0, 0, 1/4)\\
  &Co1 6i (0.5, 0, 0.24982) & Co1 6k (0.47003, 0, 0.24938)& Co2 6i (0.46960, 0, 0.24920)& Co1 12k (0.46605, 0, 0.37434)\\
  &Ge1 2c (1/3, 2/3, 0)     & Ge1 2c (1/3, 2/3, 0)       & Ge1 2c (1/3, 2/3, 0)        & Ge1 4c (1/3, 2/3, 1/4)\\
  &Ge2 2e (0, 0, 0.33710)   & Ge2 2e (0, 0, 0.33713)     & Ge2 2e (0, 0, 0.33741)      & Ge2 4e (0, 0, 0.41866)\\
  &Ge3 2d (1/3, 2/3, 1/2)   & Ge3 2d (1/3, 2/3, 1/2)     & Ge3 2d (1/3, 2/3, 1/2)      & Ge3 4d (1/3, 2/3, 0)\\
\end{tabular}
\end{ruledtabular}
\end{table}

%

\newpage

\begin{figure}[tbp]
\includegraphics[width=0.9\linewidth]{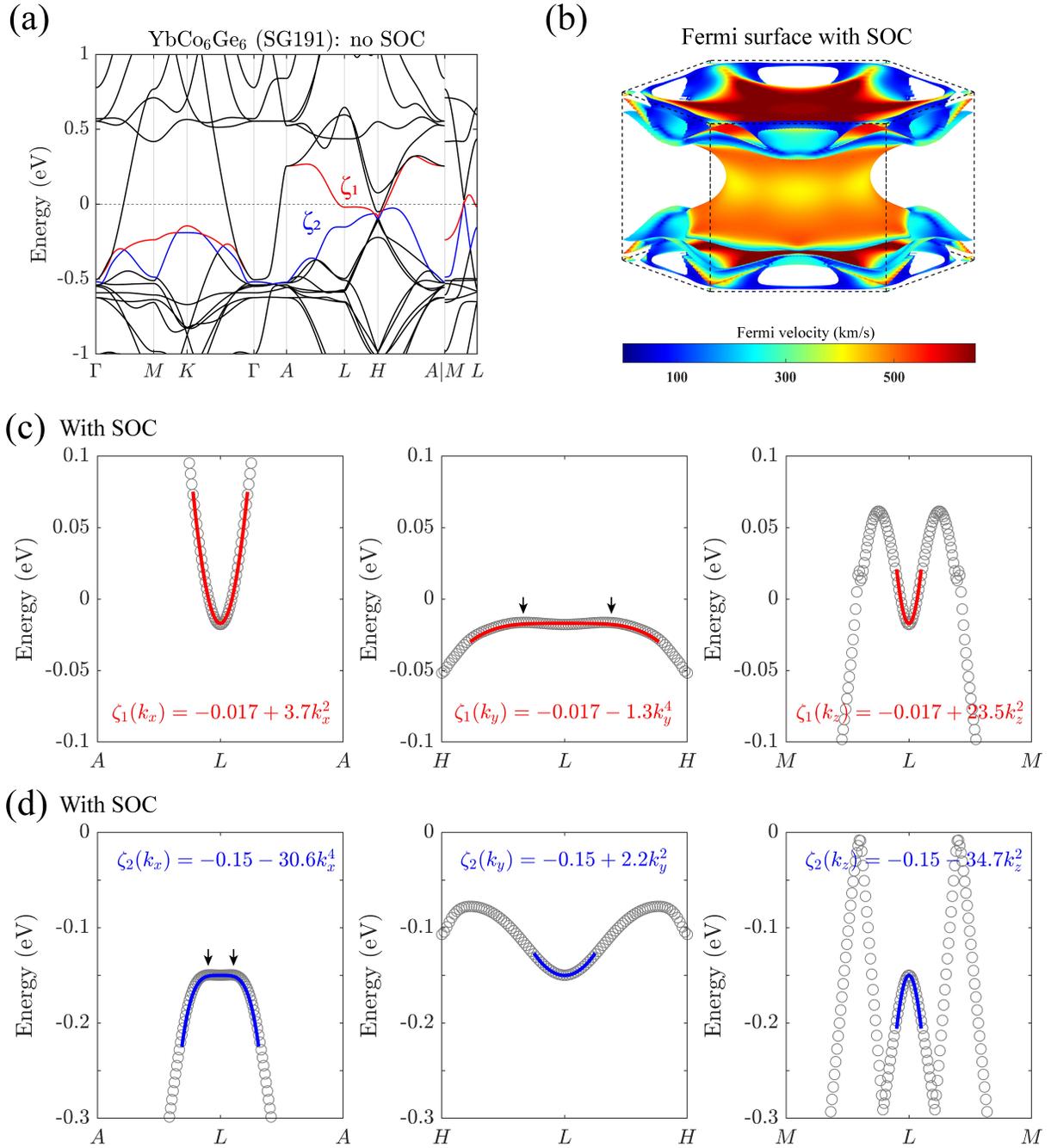}
\caption{ \label{FigSIII1} (a) Electronic band structure of the pristine YbCo$_6$Ge$_6$ ($P6/mmm$) without spin-orbit coupling (SOC).
(b) Fermi surface with SOC.
(c) and (d), polynomial fittings for the DFT dispersions of $\zeta_1$ and $\zeta_2$ along three directions ($L-A$ as $x$, $L-H$ as $y$, and $L-M$ as $z$). Grey dots are DFT data, and colored curves are fittings. The DFT bands are calculated with SOC (the full band structure is shown in the main text).
\textbf{Notice that $\zeta_1$ inherently has a $k_y^2$-dependence and $\zeta_2$ a $k_x^2$-dependence, whose coefficients are typically about 0.1 eV$\cdot$\AA$^2$.
Such quadratic terms generate two SPs on $L-H$ for $\zeta_1$ and two on $L-A$ for $\zeta_2$ (marked with black arrows), rather close to the $L$ point, both energetically and spatially, due to the smallness of the coefficients.
Merging such SPs at $L$ by ignoring the $k_y^2$ term in $\zeta_1$ and $k_x^2$ term in $\zeta_2$ fits the DFT data very well in (c) and (d).}
Thus, we ignore these terms in the main text. Besides, the resemblance between the DOS of $\zeta_1$ and $\zeta_2$ in Fig. 2(e) and those of $\epsilon_1$ and $\epsilon_2$ in Fig. 1(c)\&(d) in the main text also indicates the justification of ignoring these terms.
}
\end{figure}

\begin{figure}[tbp]
\includegraphics[width=0.9\linewidth]{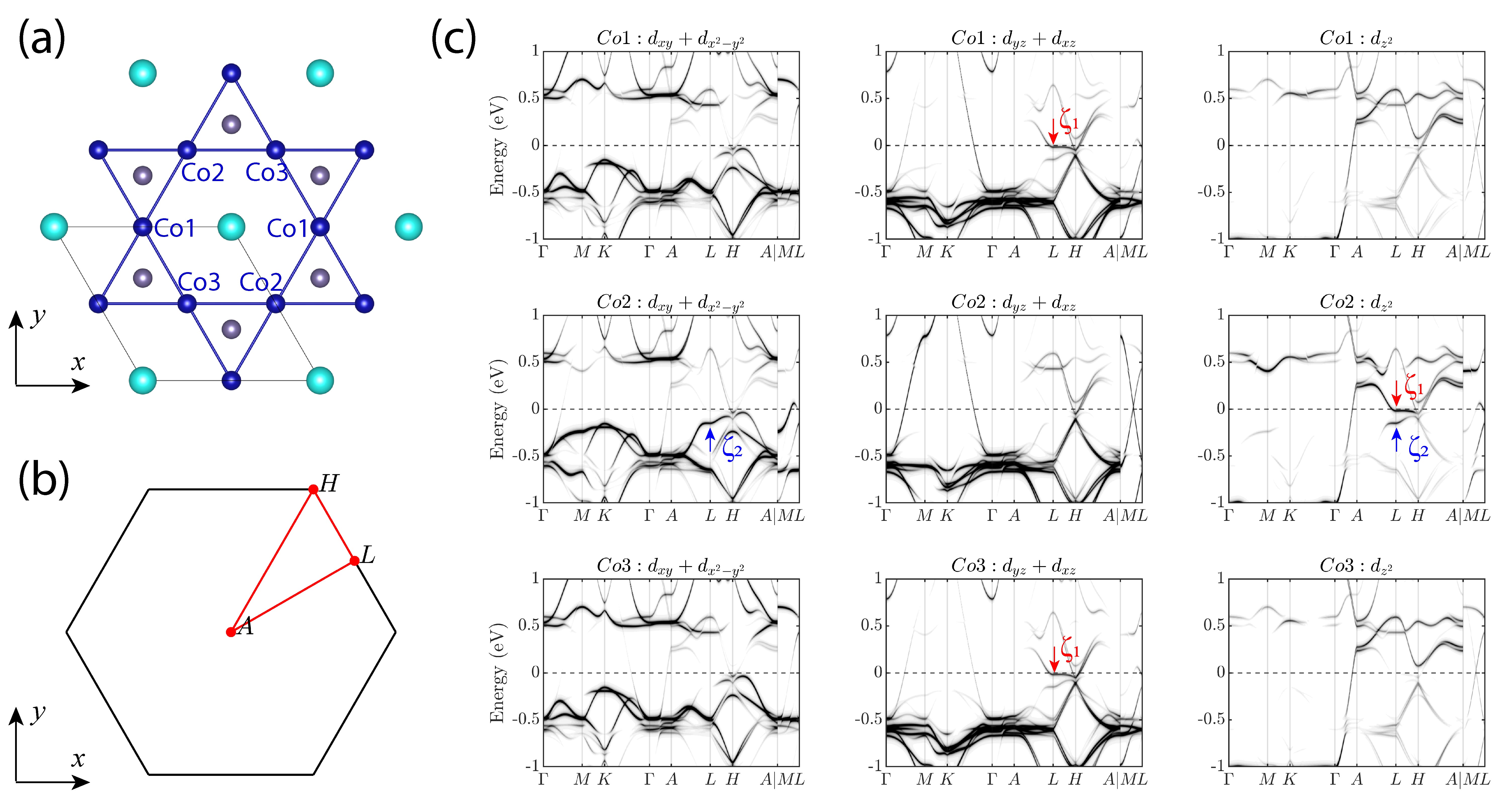}
\caption{ \label{FigSIII2} (a) shows the sublattice labels of the pristine YbCo$_6$Ge$_6$ ($P6/mmm$) where two kagome layers are overlaid. (b) depicts the $k$-paths in the $k_z=\pi$ plane. (c) presents the orbital-resolved band structures with SOC (from PBE calculations). The HOSP $\zeta_1$ is a mixed-type saddle point (SP) contributed by Co1/Co3 $d_{yz}/d_{xz}$ and Co2 $d_{z^2}$; the HOSP $\zeta_2$ is a pure-type SP dominated by Co2 $d_{xy}/d_{x^2-y^2}$ and $d_{z^2}$.
}
\end{figure}

\begin{figure}[tbp]
\includegraphics[width=0.9\linewidth]{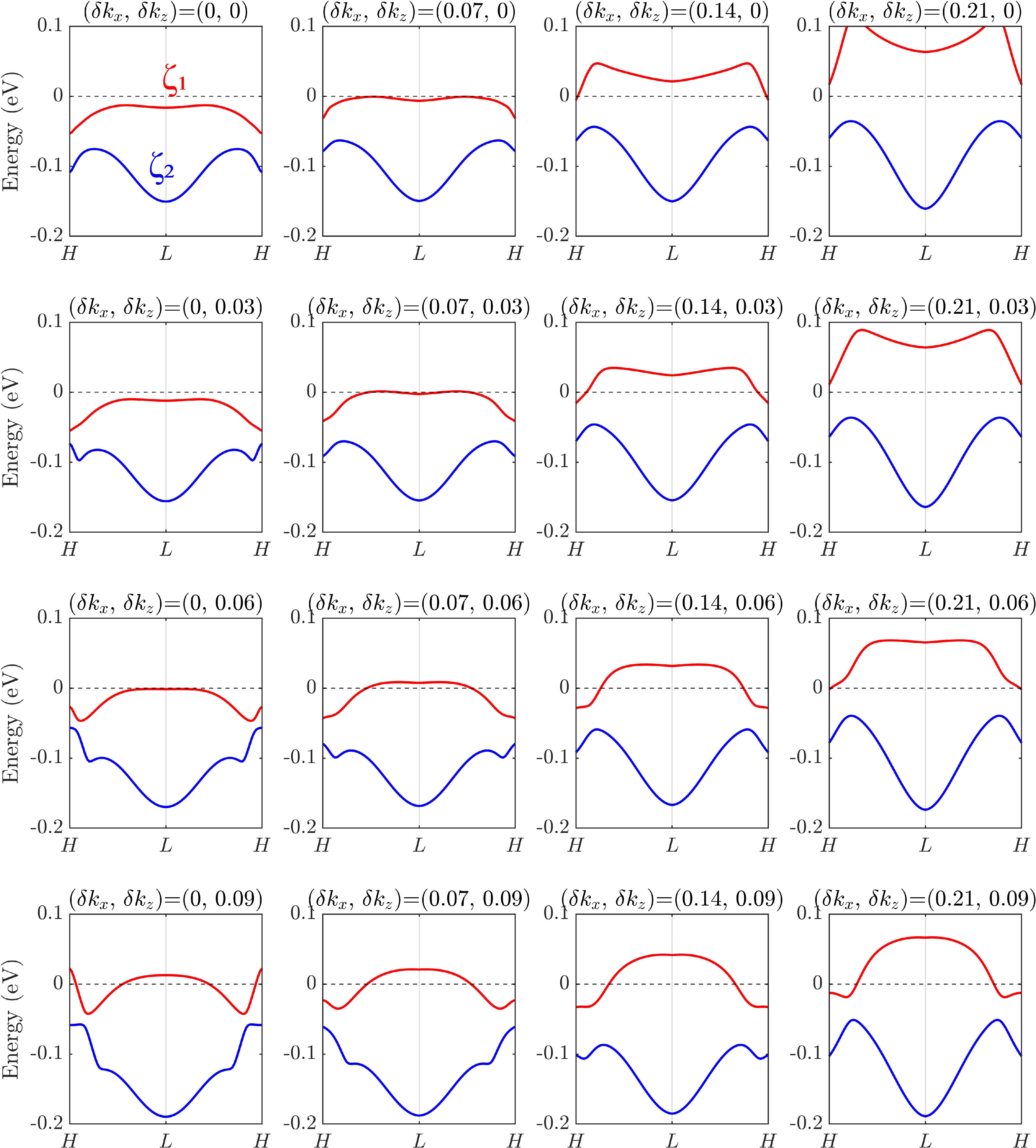}
\caption{ \label{FigSIII3} Dispersions of the HOSPs $\zeta_1$ (red) and $\zeta_2$ (blue) bands of the pristine YbCo$_6$Ge$_6$ ($P6/mmm$) along the $L-H$ line.
($\delta k_x, \delta k_z$) means the $H-L-H$ path is moved along the $k_x$ and $k_z$ direction by $\delta k_x$ and $\delta k_z$ (in unit of $\AA^{-1}$), respectively. $k_x$, $k_y$, and $k_z$ are along $L-A$, $L-H$, and $L-M$, respectively.
Compared to the $\zeta_1$ band along the original $H-L-H$ path in the upper left panel, moving the $k$-path would lead to an even flatter $\zeta_1$ band at the Fermi energy, which results in the total DOS peak at the Fermi energy Fig. 2(c) in the main text.
}
\end{figure}

\begin{figure}[tbp]
\includegraphics[width=0.85\linewidth]{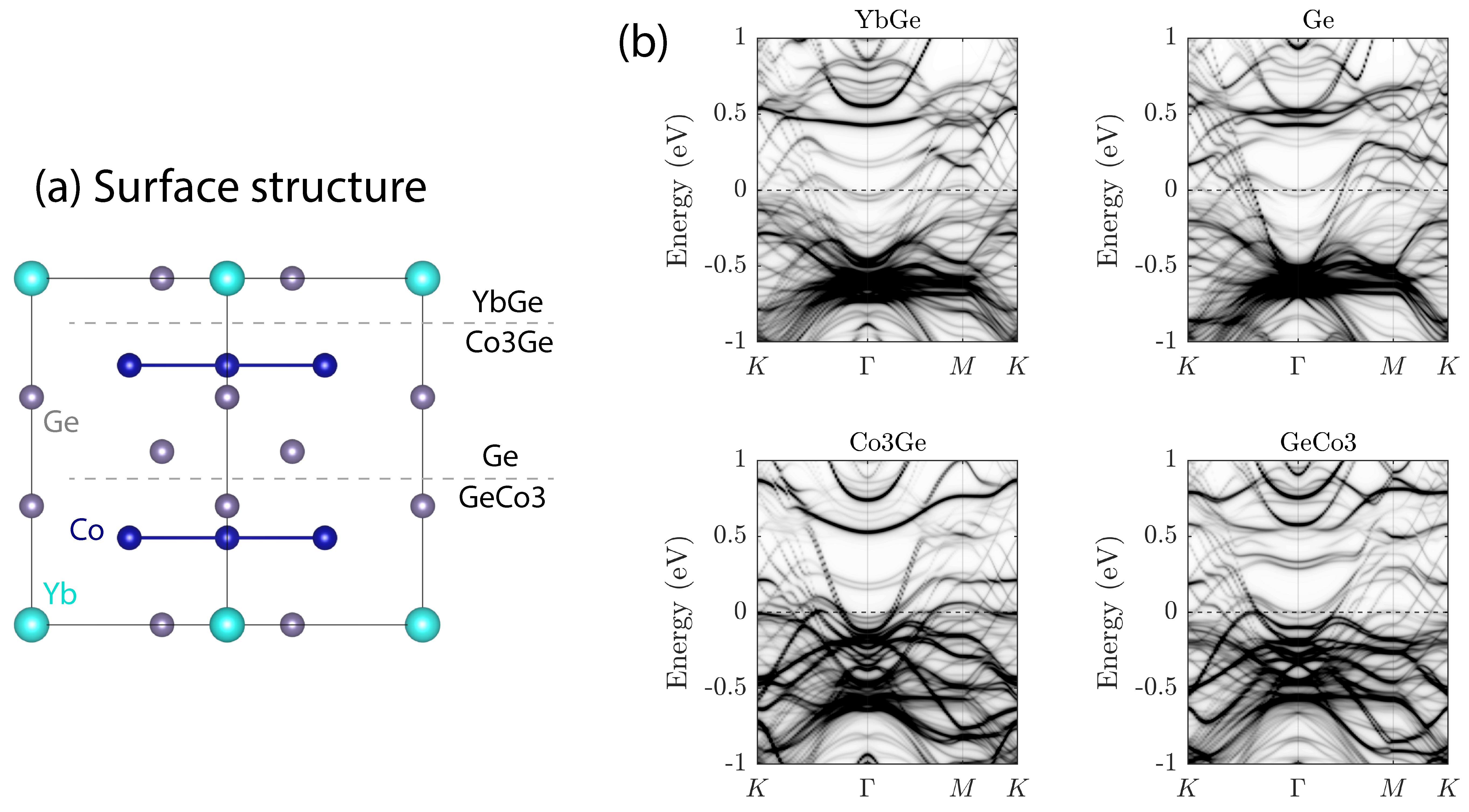}
\caption{ \label{FigSIII4} Surface states of the pristine YbCo$_6$Ge$_6$ as obtained from slab model calculations. (a) shows four possible terminations above and below the cleaving positions (grey dashed lines), i.e., YbGe, Co3Ge, Ge, and GeCo3. (b) are the corresponding surface states obtained projecting the slab band structure on the surface unit-cell thick sub slab. The slab thickness is about 6 unit cells. SOC is considered. In calculations, surfaces are fully relaxed until the remaining forces on surface atoms are less than 10 meV/\AA. Due to multiple bulk bands in the energy range, it is challenging to distinguish the non-trivial surface states from the band structures.
}
\end{figure}

\begin{figure}[tbp]
\includegraphics[width=0.9\linewidth]{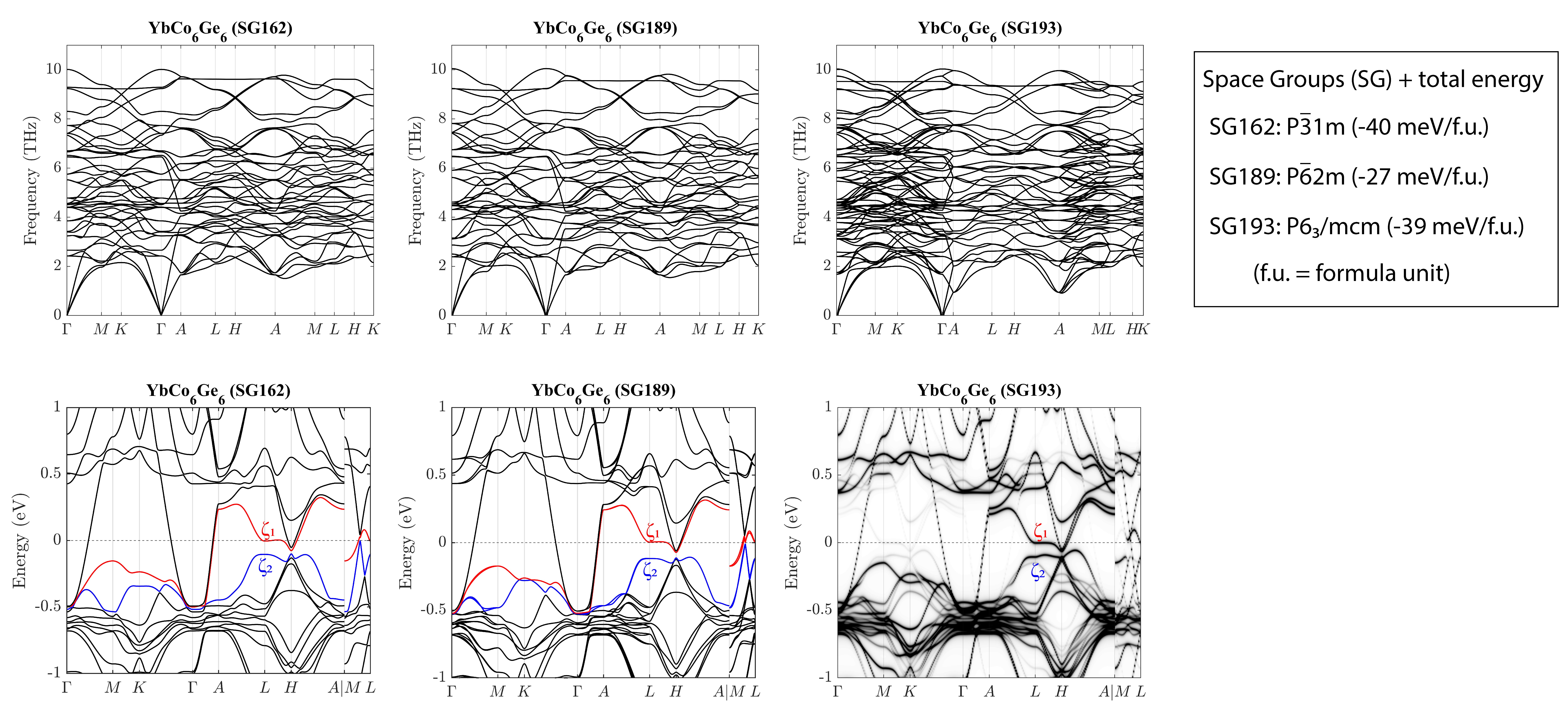}
\caption{ \label{FigSIII5} Phonon and electron band structures of YbCo$_6$Ge$_6$ in different phases, the upper row for phonon and the lower for electron.
As shown, different phases have similar influences on two HOSPs.
The upper right corner presents total energies relative to the pristine phase ($P6/mmm$). The $P\bar{3}1m$ has the lowest total energy. The $P6_3/mcm$ phase (i.e., the 1$\times$1$\times$2 CDW phase) has a similar total energy as that of the $P\bar{3}1m$ phase. 
}
\end{figure}

\begin{figure}[tbp]
\includegraphics[width=0.9\linewidth]{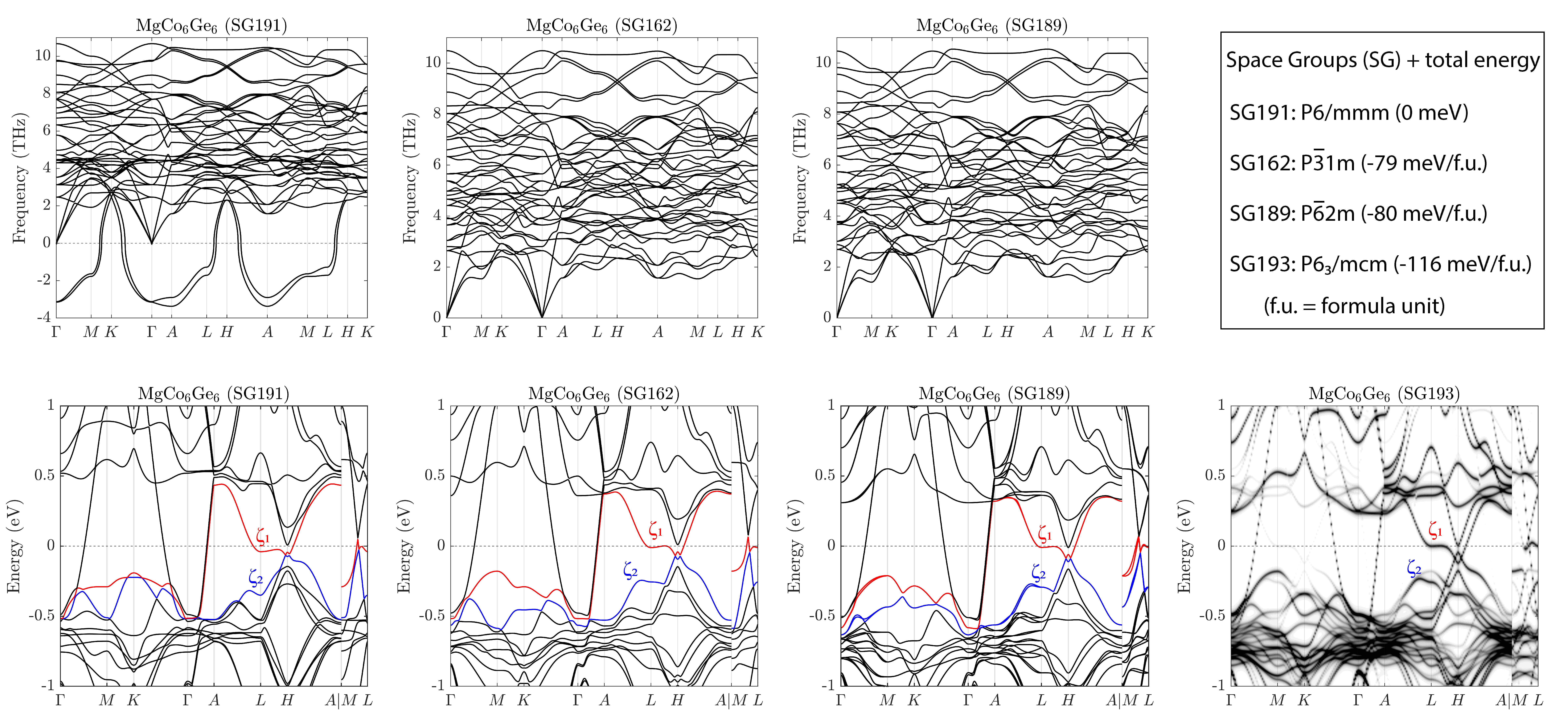}
\caption{ \label{FigSIII6} Phonon and electron band structures of MgCo$_6$Ge$_6$ in different phases, the upper row for phonon and the lower for electron.
Similar to YbCo$_6$Ge$_6$, the imaginary phonon modes at $\Gamma$ point give rise to two distorted phase $P\bar{3}1m$ (SG162) and $P\bar{6}2m$ (SG189), and the $A$ imaginary phonon modes at $A$ give rise the 1$\times$1$\times$2 CDW phase ($P6_3/mcm$, SG193). These distorted phases are all dynamically stable and energy-favored compared to the pristine phase $P6/mmm$.
As shown, different distortion phases shift the HOSP $\zeta_1$ to the Fermi energy without additional changes. For $\zeta_2$, the structural distortions turn it into a small flat band.
Total energies of different phases relative to the pristine one ($P6/mmm$) are shown in the upper right corner. Unlike YbCo$_6$Ge$_6$, the 1$\times$1$\times$2 CDW phase ($P6_3/mcm$, SG193) has the lowest total energy in MgCo$_6$Ge$_6$.
}
\end{figure}

\begin{figure}[tbp]
\includegraphics[width=0.9\linewidth]{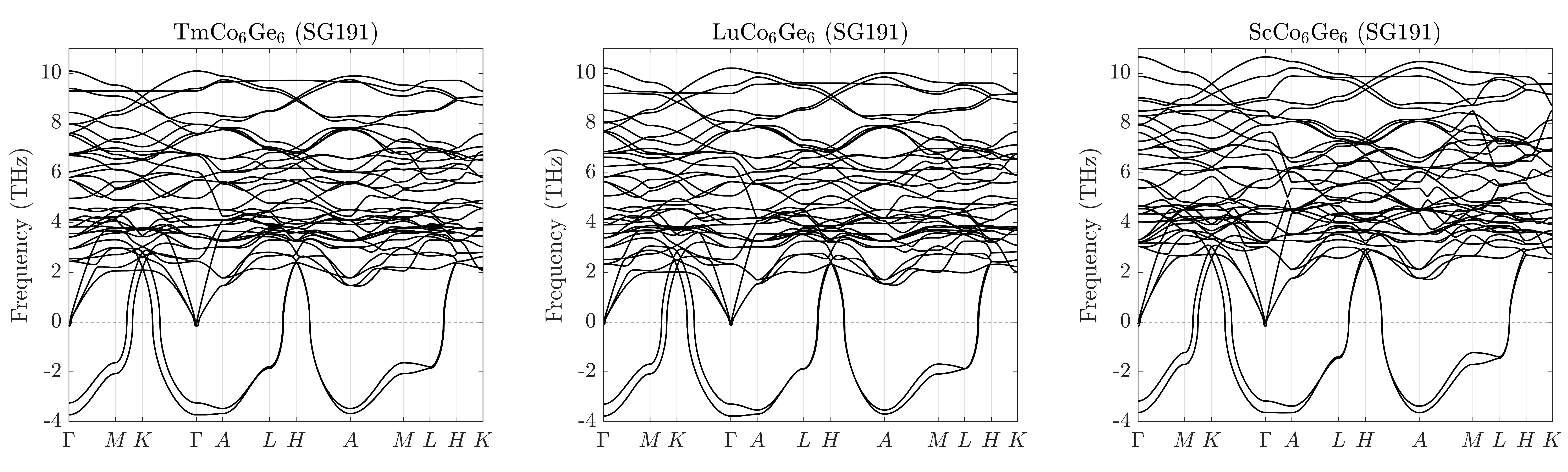}
\caption{ \label{FigSIII7} Phonon dispersions of pristine TmCo$_6$Ge$_6$, LuCo$_6$Ge$_6$, and ScCo$_6$Ge$_6$ ($P6/mmm$). The universal imaginary phonon modes at $\Gamma$ and $A$ indicate the prevalent lattice distortions in these Co-based kagome metals.
}
\end{figure}

\begin{figure}[tbp]
\includegraphics[width=0.85\linewidth]{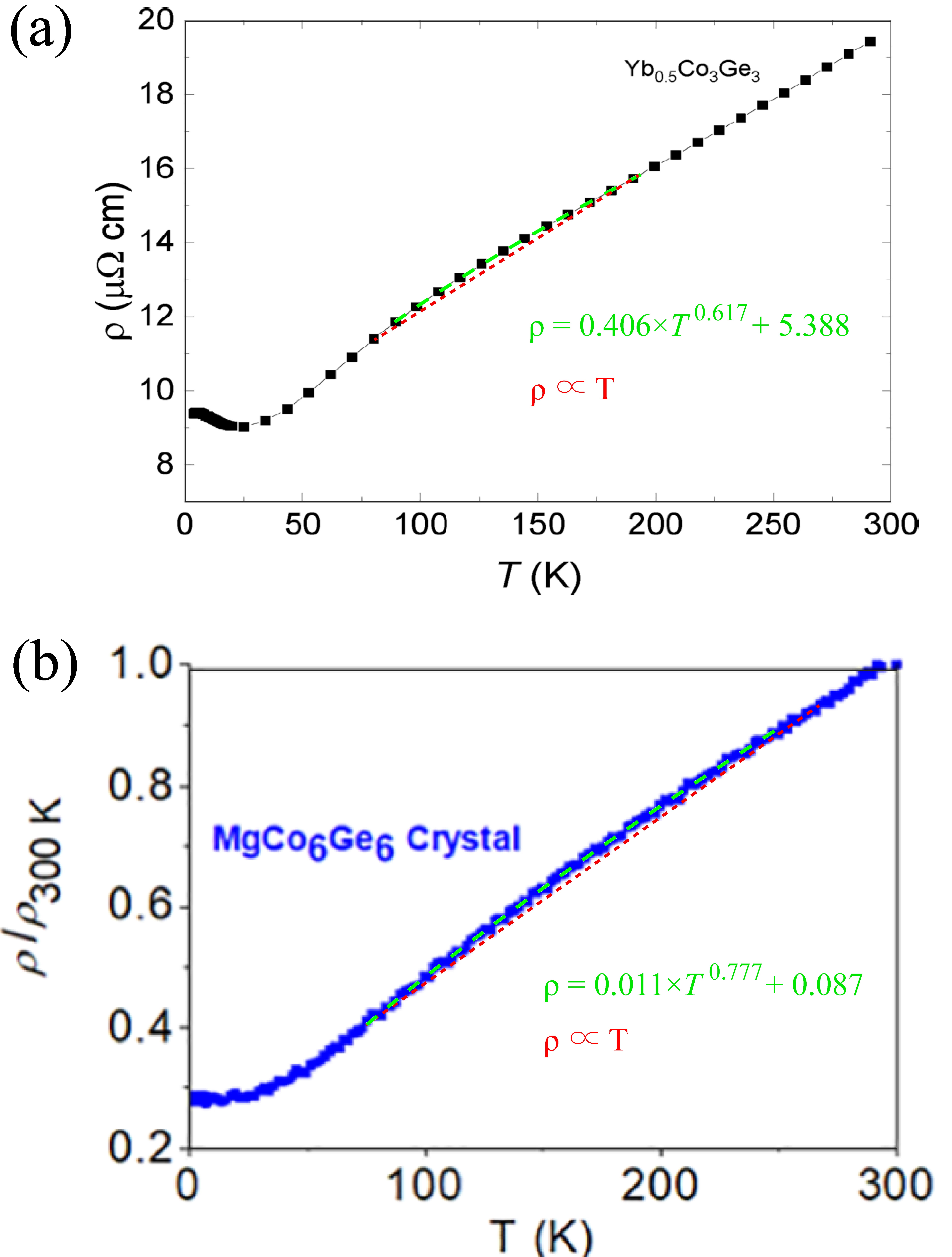}
\caption{ \label{FigSIII8} Temperature-dependent resistivity ($\rho$) of YbCo$_6$Ge$_6$ obtained with permission from Ref. \cite{weiland2020refine} (see also \cite{wang2022electronic}) and MgCo$_6$Ge$_6$ from Ref. \cite{Sinha2021twisting}. Experimental data of YbCo$_6$Ge$_6$ at temperature range (75, 200) K follow the expression $\rho = 0.406 \times T^{0.617} + 5.388$; data of MgCo$_6$Ge$_6$ at (75, 250) K show a relation $\rho/\rho_{300K} = 0.011 \times T^{0.777} + 0.087$. $T$ is temperature. Such fittings are shown in green curves. The red dash lines are linear guidelines for the eye. 
These sublinear temperature-dependent resistivities deviate from i) $T^2$-dependence of the Fermi liquid behavior, ii) $T^5$-dependence from the electron-phonon interaction at low temperature (below the Debye temperature), and iii) linear $T$-dependence at high temperature (above the Debye temperature). Such sublinear temperature dependences might arise from the close vicinity of Dirac points to saddle points near Fermi energy as proposed recently by Ref. \cite{peshcherenko2024sublinear}. Thus, we speculate the sublinear transport behavior of both kagome materials may signify the existence of HOSPs proposed in the main text.
}
\end{figure}